\documentclass[aps,letterpaper,twocolumn,preprintnumbers,floatfix,superscriptaddress,nofootinbib]{revtex4-1}

\usepackage{amsmath}
\usepackage{amsfonts}
\usepackage{amssymb}
\usepackage{graphicx, rotating}
\usepackage{epstopdf}
\usepackage{epsfig}
\usepackage{latexsym}
\usepackage{color}
\usepackage[dvipsnames]{xcolor}
\usepackage{multirow}
\usepackage{xr}
\usepackage{xfrac}
\usepackage{float}

\usepackage{xcolor,colortbl}

\usepackage{slashed}
\usepackage{hyperref}
\hypersetup{colorlinks, citecolor=bluscuro, linkcolor=bluscuro, urlcolor=bluscuro}
\definecolor{rossos}{cmyk}{0,1,1,0.55}
\definecolor{bluscuro}{rgb}{0.15, 0.2, .85}
\definecolor{bluchiaro}{cmyk}{1,.3,0.,0.1}
\definecolor{verdescuro}{rgb}{0.3,0.8,0.3}

\newcommand{\nn}{\nonumber}

\newcommand{\be}{\begin{equation}}
\newcommand{\ee}{\end{equation}}          
\newcommand{\bea}{\begin{eqnarray}}
\newcommand{\eea}{\end{eqnarray}}

\def\Re{{\rm Re\,}}
\def\Im{{\rm Im\,}}
  
 \def\Disc{\mathrm{Disc}\,}

\begin{document}
\preprint{QMUL-PH-25-06}
\widetext

\title{Cross-Section Bootstrap: Unveiling the Froissart Amplitude}

\author{Miguel Correia}
\affiliation{Department of Physics, McGill University, 3600 Rue University, Montr\'eal, H3A 2T8, QC Canada}
\author{Alessandro Georgoudis}
\affiliation{Centre for Theoretical Physics, Department of Physics and Astronomy,
Queen Mary University of London, Mile End Road, London E1 4NS, United Kingdom}
\author{Andrea L. Guerrieri}
\affiliation{CERN, Theoretical Physics Department, CH-1211 Geneva 23, Switzerland}
\affiliation{Department of Mathematics, City St. George's, University of London, \\Northampton Square, EC1V 0HB, London, UK}

\begin{abstract}
\noindent We derive a universal bound on the integrated total scattering cross-section at \emph{finite} energies, expressed in terms of a single low-energy coefficient constrained by the non-perturbative S-matrix Bootstrap. At high energies, the bound is compared with proton-proton scattering data; at low energies, with numerical bootstrap results obtained by directly maximizing the cross-section. We conjecture that the amplitude saturating the cross-section at high energies lies at a strongly-coupled corner of the allowed space of low-energy parameters. This universal amplitude exhibits a rising total cross-section, a shrinking elastic differential cross-section with multiple diffractive minima, and a surprisingly rich spectrum of resonances aligning along Regge trajectories, including Pomeron-like and daughter trajectories, as well as unusual ``singular" trajectories in the forward limit which appear deeply interconnected with Froissart growth.  Remarkably, the eikonal representation reveals that the scattering is localized within an annular region that slowly expands with energy, challenging the traditional ``disk" diffraction picture. Our results open the door to theoretical and phenomenological studies of \emph{soft} high-energy hadronic scattering via the S-matrix Bootstrap.
\end{abstract}

\maketitle
\medskip

\section{Introduction}

\noindent Quantum Chromodynamics (QCD), the theory of strong interactions, governs a vast array of physical phenomena across different energy scales, each presenting unique theoretical challenges. At high energies, where all relevant scales are large, asymptotic freedom ensures that QCD becomes weakly coupled. In this regime, perturbation theory (PT) provides a systematic and predictive framework, allowing for first-principles calculations \cite{Gross:2022hyw}.

At low energies, however, QCD becomes strongly coupled, with quarks and gluons confined into hadrons, rendering analytical solutions intractable. In this regime, the dynamics of light hadrons---pions and other pseudo-Goldstone bosons---are successfully described by a combination of effective field theories (EFTs), such as the chiral Lagrangian, numerical techniques like lattice QCD, and modern bootstrap or dispersive methods. Together, these approaches capture the non-perturbative dynamics of QCD in the infrared.

The situation becomes dramatically more challenging in the case of forward scattering at high energies, where the momentum transfer is small: the so-called \emph{soft} regime of QCD. In this regime, neither PT nor EFT methods apply, leaving few options beyond general principles such as unitarity, analyticity, and crossing symmetry, which together form the foundation of the \emph{bootstrap} approach.

The bootstrap philosophy, which aims to determine physical observables from consistency conditions alone, occupies a central position in modern theoretical physics. Its roots trace back to the 1960s, precisely in the context of strong interactions \cite{Chew:1962eu, chew1966analytic}, when experimental observations of the \emph{rising} total cross-section in proton-proton scattering at high energies \cite{Amaldi:2015jhq} suggested that QCD might saturate the celebrated Froissart-Martin bound \cite{PhysRev.123.1053, martinbound}:
\begin{equation} 
\label{eq:Froissart} 
\sigma_\text{tot}(s) \leq \frac{4 \pi}{t_0} \log^2 \Big({s \over s_0} \Big), \qquad s \to \infty, \end{equation}
for some $s_0 > 0$, where $\sigma_\text{tot}$ is the total scattering cross-section and $t_0 = (2 m_\pi)^2$ is the two-pion production threshold. 

The discovery that hadron masses and spins align along approximately linear \emph{Regge trajectories} led to the development of Regge theory. This framework embodied the bootstrap philosophy and provided the first semi-quantitative descriptions of QCD in the soft regime. Notably, it predicted the existence of the conjectured \emph{Pomeron} trajectory, thought to underlie the growth of the total cross-section. Yet, despite these successes, Regge theory ultimately evolved into a largely phenomenological tool with limited predictive power.

Interestingly, modern measurements at the LHC (\(\sqrt{s} = 13\,\text{TeV}\)) are consistent with Froissart-like growth of the proton-proton total cross-section, \(\sigma_\text{tot}(s) \simeq c \log^2 s\)~\cite{COMPETE:2002jcr, TOTEM:2017asr,ParticleDataGroup:2024cfk}. However, the measured coefficient \(c\) is two orders of magnitude smaller than the theoretical bound \(4\pi / t_0 = \pi/m_\pi^2\) from \eqref{eq:Froissart}. This discrepancy suggests that we are not yet in the true asymptotic regime~\cite{Grau:2012wy,Block:2012nj, Dymarsky:2014rba, ParticleDataGroup:2024cfk}, but rather in a transitional regime where multi-Pomeron exchanges, associated with Regge cuts, are beginning to play a role~\cite{Ostapchenko:2010gt,Khoze:2014aca,ParticleDataGroup:2024cfk}. Additional explanations include the softness of \(\pi\pi\) exchange, which may suppress its contribution in the Regge limit~\cite{ANSELM1972487,stephanov,Giordano:2013iga}, and the fact that the Froissart bound is derived under limited assumptions, making it unlikely to be optimal~\cite{Martin:2008xb,Kupsch:2008hq,Martin:2013xcr}.

These considerations, combined with the fact that soft interactions, which dominate the ``sea'' from which new physics must be extracted at the LHC~\cite{ParticleDataGroup:2024cfk}, remain poorly understood, underscore the need for a systematic framework to describe hadronic scattering in this regime. This Letter argues that the modern, non-perturbative S-matrix
Bootstrap~\cite{Paulos:2016but,Paulos:2017fhb,Doroud:2018szp,He:2018uxa,Cordova:2018uop,Guerrieri:2018uew,Homrich:2019cbt,Karateev:2019ymz,EliasMiro:2019kyf,Paulos:2018fym,Bercini:2019vme,Cordova:2019lot,Kruczenski:2020ujw,Guerrieri:2020kcs,Bose:2020cod,Bose:2020shm,Guerrieri:2020bto,Hebbar:2020ukp,Sinha:2020win,Guerrieri:2021ivu,Correia:2021etg,Tourkine:2021fqh,Karateev:2022jdb,EliasMiro:2021nul,He:2021eqn,Guerrieri:2021tak,Chen:2022nym,EliasMiro:2022xaa,Miro:2022cbk, Guerrieri:2022sod,Haring:2022sdp,Albert:2022oes,Correia:2022dyp,PhysRevD.110.025012,Acanfora:2023axz,Albert:2023jtd,Antunes:2023irg,Albert:2023seb,Guerrieri:2023qbg,Cordova:2023wjp,EliasMiro:2023fqi,Tourkine:2023xtu,Gaikwad:2023hof,Bhat:2023puy,He:2024nwd,Copetti:2024rqj,Copetti:2024dcz,Albert:2024yap,Guerrieri:2024ckc,Guerrieri:2024jkn,Gumus:2024lmj,Cordova:2025bah,He:2025gws}, 
especially when combined with analytical methods, can offer precisely such a framework.

We begin by addressing the fundamental limitation of the Froissart bound: its purely asymptotic nature prevents direct comparison with experimental data. Pointwise bounds at finite energies are generally ruled out due to the distributional character of $\sigma_\text{tot}(s)$.\footnote{An infinite accumulation of higher-spin resonances at a single point $s_0$ can lead to $\sigma_\text{tot}(s \to s_0) \to \infty$; see Appendix~\ref{sec:divcross}.} However, integrated quantities can still admit meaningful bounds. Notably, Yndurain~\cite{Yndurain:1972ix} demonstrated that the integrated cross-section for the scattering of identical scalar particles of mass $m$,
\begin{equation} 
\label{eq:integrated_cross_section} 
\bar{\sigma}_\text{tot}(s) \equiv \frac{1}{16\pi} \int\limits_{4m^2}^s \frac{s' - 4m^2}{s - 4m^2} \,\sigma_\text{tot}(s') \, ds' 
\end{equation} 
is bounded from above under the assumption that the spin-2 (or $D$-wave) partial wave scattering length is finite, and known.

Here we remove this key assumption and derive universal quantitative bounds on the integrated total cross-section for identical scalar particles in any number of dimensions. Our bounds follow from the crucial requirement of crossing symmetry and non-linear unitarity enabled by the modern, non-perturbative S-matrix bootstrap \cite{Paper3}.

\begin{figure}
    \centering
    \includegraphics[width=\linewidth]{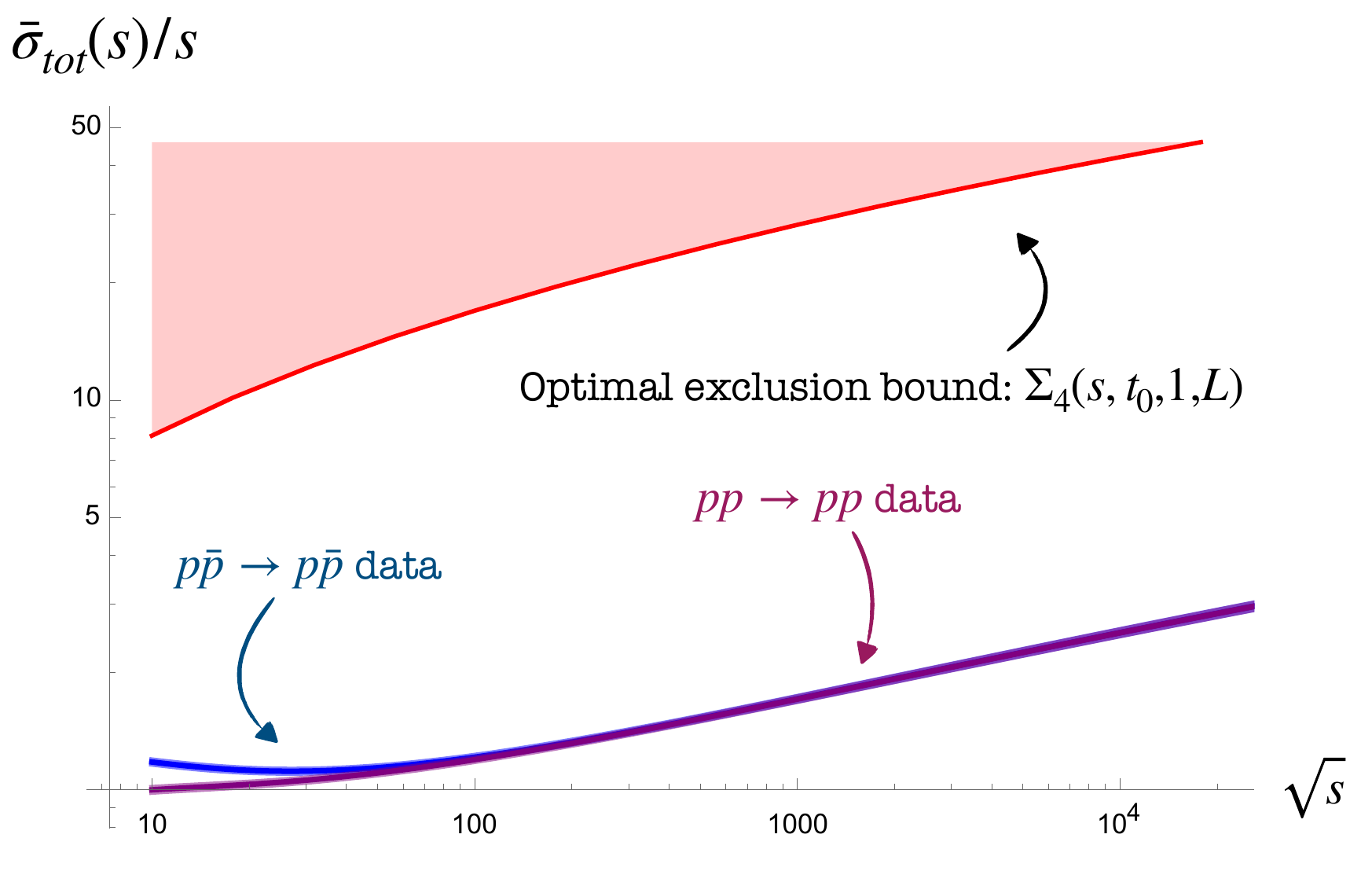}
    \caption{The integrated cross-section \eqref{eq:integrated_cross_section} for $pp$ and $p\bar p$ scattering is plotted respectively in purple and blue  using the phenomenological fits discussed in~\cite{ParticleDataGroup:2016lqr}, and expressed in dimensionless units (setting $1\text{GeV}=1$).  The thickness of the curves include $3\sigma$ deviations from the best fit. The red excluded region is the optimal exclusion bound \eqref{eq:sigmatotanalytic} in the same units. 
    }
    \label{fig:LHC_vs_us}
\end{figure}

The scattering of identical scalars provides an ideal model to explore the universal properties of high-energy dynamics, as spin effects become negligible in this regime. This simplification makes it meaningful to compare our bounds with experimental data for proton-proton and proton–antiproton scattering as shown in Fig.~\ref{fig:LHC_vs_us}. As the figure illustrates, the Bootstrap exclusion bound lies within a factor of ten of the experimental data. In doing so, we assume $m \sim 1 \,\text{GeV}$ as the relevant range of interaction in \( pp \) and \( p\bar{p} \) scattering at high energies \cite{ANSELM1972487}.

Although the integrated cross-section \eqref{eq:integrated_cross_section} diverges in the high-energy limit $s\to\infty$, we conjecture that the scattering amplitude which saturates this quantity admits a well-defined, finite limiting form. This asymptotic function, which we refer to as the \emph{Froissart amplitude}, has emerged in several bootstrap contexts~\cite{Paper3,Guerrieri:2021tak,Guerrieri:2023qbg, Bhat:2023puy}. Here we determine this amplitude with unprecedented numerical precision, allowing us to study it for the first time in the physical scattering region. It displays several remarkable features such as  a rising total cross-section, a shrinking diffractive cone, and an intricate spectrum of resonances aligning along Regge trajectories, which are distinctive qualitative features of hadronic scattering. These results suggest that the S-matrix Bootstrap can be used to study QCD in the soft regime.

The Letter is organized as follows. Section~\ref{sec:anbound} derives the analytical bound on the cross-section at finite energy. In Section~\ref{sec:numerical_s_matrix_bootstrap}, we introduce the modern S-matrix Bootstrap framework and present the numerical bounds it yields. Section~\ref{sec:Froissart_amplitude} is dedicated to exploring the physics of the Froissart amplitude. We conclude in Section~\ref{sec:conclusion} with a discussion and outlook.

\section{Bound on cross-section at finite energy in any dimension}
\label{sec:anbound}

\noindent In this section, we report the main steps in the derivation of the analytic bound. The strategy we present builds up on the classical works \cite{PhysRev.123.1053, Yndurain:1972ix}. For simplicity, we consider the scattering of identical scalars,\footnote{Generalizing our argument for non-identical spinning particles should be straightforward.} where we denote $T(s,t)$ as the $2 \to 2$ scattering amplitude, with $s$ and $t$ the usual Mandelstam variables. 

We will need the S-matrix element for the scattering of angular momentum $\ell$ eigenstates,
\begin{equation}
\label{eq:Sl}
    S_\ell(s) = 1+ i\mathbf{N}_d(s) \int\limits_{-1}^1 dz \, (1-z^2)^{\tfrac{d-4}{2}}\mathcal{P}^{(d)}_\ell(z) \, T(s,t(z)),
\end{equation}
also known as the \emph{partial wave amplitude}, where $\mathbf{N}_d(s)=\mathcal{N}_d \,\frac{(s-4m^2)^{\tfrac{d-3}{2}}}{2\sqrt{s}}$, and $\mathcal{N}_d=\frac{(16\pi)^{1-\tfrac{d}{2}}}{\Gamma(\tfrac{d}{2}-1)}$.
$\mathcal{P}^{(d)}_\ell(z)$ can be expressed in terms of Gegenbauer polynomials $\mathcal{P}_\ell^{(d)}(z)=C^{(d-3)/2}_\ell(z)/C^{(d-3)/2}_\ell(1)$
normalized such that $\mathcal{P}^{(d)}_\ell(1) = 1$, and $z = 1 + {2 t \over s- 4m^2}$ is the cosine of the scattering angle in the $s$-channel. 
Unitarity in this basis takes the simple form $|S_\ell(s)|^2\leq 1$
for all $s \geq 4 m^2$, where $m = 1$ henceforth.
The amplitude can be expanded in partial waves
\begin{equation}\label{eq:T_partial_waves}
    T(s,t) = {i \sqrt{s} \over (s-4)^{\tfrac{d-3}{2}}} \sum_{\ell = 0}^\infty n_\ell^{(d)} \, \mathcal{P}^{(d)}_\ell(z) \, \big(1 - S_\ell(s)\big),
\end{equation}
with $n_\ell^{(d)}=\frac{(4\pi)^{\tfrac{d}{2}}(d{+}2\ell{-}3)\Gamma(d{+}\ell{-}3)}{\pi \Gamma(\ell+1)\Gamma(\tfrac{d}{2}-1)}$. Via the optical theorem,
\be
\label{eq:optical}
\sigma_\text{tot}(s)=\frac{\Im T(s,0)}{\sqrt{s(s-4)}},
\ee
the total cross-section admits a simple partial wave expansion
\be
\label{eq:cross-section}
\sigma_\text{tot}(s)=\sum_{\ell=0}^\infty n_\ell^{(d)}\frac{1-\Re S_\ell(s)}{(s-4)^{d/2-1}}.
\ee
Free propagation implies $S_\ell(s)\to 1$, causing the total cross-section $\sigma_\text{tot}(s)$ to vanish. To maximize \eqref{eq:cross-section}, we must therefore minimize $\Re S_\ell(s)$. The unitarity condition sets the minimum at $S_\ell(s)=-1$. However, in the presence of a mass gap, interactions have finite range, requiring $S_\ell(s)\to 1$ at large angular momentum $\ell$, which effectively regulates the partial wave sum in \eqref{eq:cross-section} and makes the cross-section finite.
Next, we introduce the integrated cross-section in $d$ dimensions\footnote{The kernel is introduced to soften the threshold behavior $s \to 4$ appearing in \eqref{eq:T_partial_waves}.}
\be
\label{eq:sigmabartot_in_d_dimensions}
\bar\sigma_\text{tot}(s)=\mathcal{N}_d \int\limits_4^s ds^\prime \left(\frac{s^\prime-4}{s-4}\right)^{\tfrac{d}{2}-1}\sigma_\text{tot}(s^\prime).
\ee
We plug \eqref{eq:cross-section} into \eqref{eq:sigmabartot_in_d_dimensions} and separate the sum as follows

\bea
\label{eq:sigmasplit}
&&\!\bar{\sigma}_\text{tot}(s){=} \!\int\limits_4^s \! ds' \!\left(\sum_{\ell=0}^{L-2}+\sum_{\ell=L}^\infty\right) \!{\mathcal{N}_d n_\ell^{(d)}\big(1 {-} \Re S_l(s^\prime)\big) \over (s - 4)^{\tfrac{d}{2}-1}}\leq \nn\\
&&\!\leq  \frac{\mathcal{N}_d}{(s-4)^{\tfrac{d}{2}-2}} \frac{2^{d}\pi^{\tfrac{d}{2}-1}}{\Gamma(d/2)}(L-1)_{d-2}+\nn\\
&&\!+{\mathcal{N}_d \over (s - 4)^{\tfrac{d}{2}-1}}\int\limits_4^s \! ds' \sum_{\ell=L}^\infty { n_\ell^{(d)}\big(1 - \mathrm{Re} S_l(s^\prime)\big) }
\eea
for some finite arbitrary even $L$.\footnote{Note that due to Bose symmetry of the amplitude $T(s,t) = T(s,u)$ the odd spins $\ell$ do not contribute. This follows from the relation $z_s(u) = - z_s(t)$ and $\mathcal{P}^{(d)}_\ell(-z) = (-1)^\ell \mathcal{P}^{(d)}_\ell(z)$. Which for odd $\ell$ implies that the integral in \eqref{eq:Sl} vanishes and $S_{\ell = \text{odd}} = 1$.} 
We set $\Re S_\ell(s)=-1$ for $\ell < L$.
To bound the sum over the high spins $\ell \geq L$ we will relate this term to the the $k$-th derivative of the amplitude at $s=2-t_0/2$ for any $t_0 \in [0,4)$ 
\be
\label{eq:c2_bare}
\frac{2}{\mathcal{N}_d} c_{2k}(t_0)=\frac{1}{2^k (2k)!}\frac{\partial^{2k}}{\partial s^{2k}}T(s,t_0)|_{s\to 2-\tfrac{t_0}{2}}.
\ee 
Assuming polynomial boundedness $T(s,t\leq t_0)\lesssim s^N$ for $s\to\infty$,
it can be shown that 
this quantity admits a compact \emph{positive dispersive relation}\footnote{We use Cauchy's formula to write $c_k(t_0)$ in \eqref{eq:c2_bare} as a contour integral. We then blow up the contour and  make use of the Regge boundedness to drop the contribution at infinity. There will be branch cuts in the $s$- and $u$-channels starting from the $2$-particle threshold. Finally we use $T(s,t) = T(u,t)$ to re-express the $u$-channel cut in terms of the $s$-channel cut.} \cite{Adams:2006sv}
\be
\label{eq:c2k}
\frac{2}{\mathcal{N}_d} c_{2k}(t_0)=\frac{2^{1-k}}{\pi}\int\limits_4^\infty ds \frac{\Im T(s,t_0)}{(s-2+\tfrac{t_0}{2})^{2k+1}}\geq 0.
\ee
We may now plug the partial wave expansion \eqref{eq:T_partial_waves} into \eqref{eq:c2k} which expresses $c_{2k}(t_0)$ in terms of a manifestly positive kernel, and after a series of simple inequalities we get
\be
c_{2k}(t_0)\geq \frac{\mathcal{N}_d}{\pi 2^{k}}\mathcal{K}_L^{(d)}(s,t_0)\int\limits_4^s ds^\prime \sum_{\ell=L}^\infty n_\ell^{(d)}(1-\Re S_\ell(s^\prime)),
\ee
where $\mathcal{K}_\ell^{(d)}(s,t_0)=\mathcal{P}_\ell^{(d)}(1{+}\tfrac{2 t_0}{s-4})\frac{\sqrt{s}(s-4)^{\tfrac{3-d}{2}}}{(s-2+\tfrac{t_0}{2})^{2k+1}}$
which allows us to bound the high-spin $\ell$ contribution to \eqref{eq:sigmasplit} once we fix the value of $c_{2k}(t_0)$.
This yields the following strict upper bound on the cross-section
\bea
\label{eq:sigmatotanalytic}
&&(s-4)^{\tfrac{d}{2}-2}\,\bar\sigma_\text{tot}(s)\leq \\
&&\leq (s-4)^{\tfrac{d}{2}-2}\,\Sigma_d(s,t_0,k,L)=\nn\\
&&= 
\frac{\mathcal{N}_d 2^{d}\pi^{\tfrac{d}{2}-1}}{\Gamma(\tfrac{d}{2})}(L-1)_{d-2}+\frac{2^k \pi \, c_{2k}(t_0)}{(s-4) \,\mathcal{K}_L^{(d)}(s,t_0)},\nn
\eea
This bound is valid for any integer $k$, even integer $L$ and $t_0 \in (0,4)$, and if we also set $c_{2k}(t_0)\equiv \max c_{2k}(t_0)$, then this bound is valid for any theory of gapped scalars in $d$-dimensions, assuming a polynomial bound on the amplitude of the form $|T(s,t)|<s^{2k}$. 

In the limit $s\to\infty$ the bound function $\Sigma_d$ simplifies and it can be minimized analytically finding that $L^*\simeq \sqrt{s \over 4 t_0}\log s$. The optimal asymptotic bound is independent of $c_{2k}(t_0)$ and is given by (see Appendix \ref{sec:Analytical_bound} for the derivation)
\be
\label{eq:froissart_in_d_dims}
\bar{\sigma}_\text{tot}(s) \leq {4 \pi^{{d \over 2} - 1} \over \Gamma({d \over 2}) \, t_0^{{d \over 2} - 1}  } s \log^{d-2}s, \;\;\;\;\;\; s \to \infty \,.
\ee
In four dimensions it reduces to the Froissart bound \eqref{eq:Froissart} reproducing the results of \cite{Common:1970as,Yndurain:1972ix,Martin:2013xcr}. Note that in writing \eqref{eq:froissart_in_d_dims}, we set the optimal choice $k=1$ and $t_0 \to4$. This shows that dispersion relations admit at most two subtractions in any dimension.

As shown in the next section, one cannot set \( t_0 = 4 \) exactly, given that the coefficient \( c_2(t_0 \to 4) \) is unbounded due to the emergence of a spin-2 bound state at threshold. In this limit, \( c_2(4m^2) \) coincides with the \( D \)-wave scattering length, which Yndurain~\cite{Yndurain:1972ix} crucially assumed to be finite. 

We will now determine $\max c_2(t_0)$ using the S-matrix Bootstrap and find the optimal bound \eqref{eq:sigmatotanalytic} on the integrated cross-section $\bar{\sigma}_\text{tot}(s)$ at finite $s$.\footnote{It is worth noticing that \eqref{eq:sigmatotanalytic} can be used to derive rigorous bounds for the scattering of massless whenever a non-perturbative S-matrix exists. E.g. in the case of the scattering of dilatons in $d=4$ the bound reads $\bar\sigma_\text{tot}(s)\leq \Delta a/f^4 s^2$, where $\Delta a$ is the anomaly coefficient, and $f$ the Goldstone decay constant. Trivially, this also implies $\Delta a > 0$ because the cross-section is positive. See Appendix \ref{sec:Analytical_bound} for details. }

 \section{Numerical S-matrix bootstrap}
\label{sec:numerical_s_matrix_bootstrap}

\noindent Here and henceforth we set $d=4$. Our numerical setup is the one described in \cite{Guerrieri:2024jkn, EliasMiro:2022xaa}, and first introduced in \cite{Paper3}. 
We expand the non-perturbative $2\to 2$ scattering amplitude on a \emph{multi-foliation basis} $T(s,t)=\sum_{p \in P} \mathcal{F}^N_p(s,t)$, with
\be
\label{eq:foliation}
\mathcal{F}^N_p(s,t)=\sum_{a{+}b\leq N}\alpha_{(a,b)}(\rho_s^a \rho_t^b{+}\rho_s^a \rho_u^b{+}\rho_t^a\rho_u^b)
\ee
where $\rho_s=(\sqrt{p-4}-\sqrt{4-s})/(\sqrt{p-4}+\sqrt{4-s})$  maps the cut plane to the unit disk with center at $s=8-p$. It is hard to guess a priori what is the optimal set of points $p \in P$ that will guarantee optimal convergence. A good practice is to distribute $p$ in correspondence of the resonances, as discussed in \cite{Gaikwad:2023hof}. See Appendix \ref{sec:appendix_numerics_details} for a summary of the parameters used to obtain the numerical results of this paper.\footnote{An alternative strategy is to be
agnostic to the spectrum and use a more democratic pure \emph{wavelet ansatz} \cite{EliasMiro:2022xaa}.}

Crucially, the expansion \eqref{eq:foliation} implements maximal analyticity and crossing-symmetry in all variables $s$, $t$ and $u = 4 -s - t$. The free parameters \( \alpha_{(a,b)} \) become constrained upon imposing unitarity, \( |S_\ell(s)|^2 \leq 1 \), which is enforced numerically for all even spins \( \ell \leq L_\text{max} \) through the positive semi-definite condition \cite{Paper3}
\be
\begin{pmatrix}
1-\Im S_\ell(s) && \Re S_\ell(s) \\
\Re S_\ell(s) && 1+\Im S_\ell(s)
\end{pmatrix}\succeq 0,
\ee
where the partial waves are determined via \eqref{eq:Sl}. The remaining partial waves, with $\ell>L_\text{max}$, are constrained via the additional \emph{improved positivity constraints} \cite{EliasMiro:2022xaa, Guerrieri:2022sod} imposed for $s>4$ and $0\leq t<4$
\be
\Im T(s,t)-\sqrt{\tfrac{s}{s-4}}\sum_{\ell=0}^{L_\text{max}}n_\ell^{(4)} (1-\Re S_\ell(s))\mathcal{P}^{(4)}_\ell(z)\geq 0.
\label{eq:improved_pos_constraints}
\ee
This condition significantly stabilizes the numerics in $L_\text{max}$ at the value of $L_\text{max} = 18$ we consider. Numerical convergence is achieved by progressively increasing $N$, which controls the total number of parameters $\alpha_{(a,b)}$ in our expansion \eqref{eq:foliation}. We solve the bootstrap optimization problem via \texttt{SDPB} \cite{Landry:2019qug}.

\subsection{Bound on the Integrated Cross-Section}
\label{sec:bound_on_integrated}

\begin{figure}
    \centering
\includegraphics[width=\linewidth]{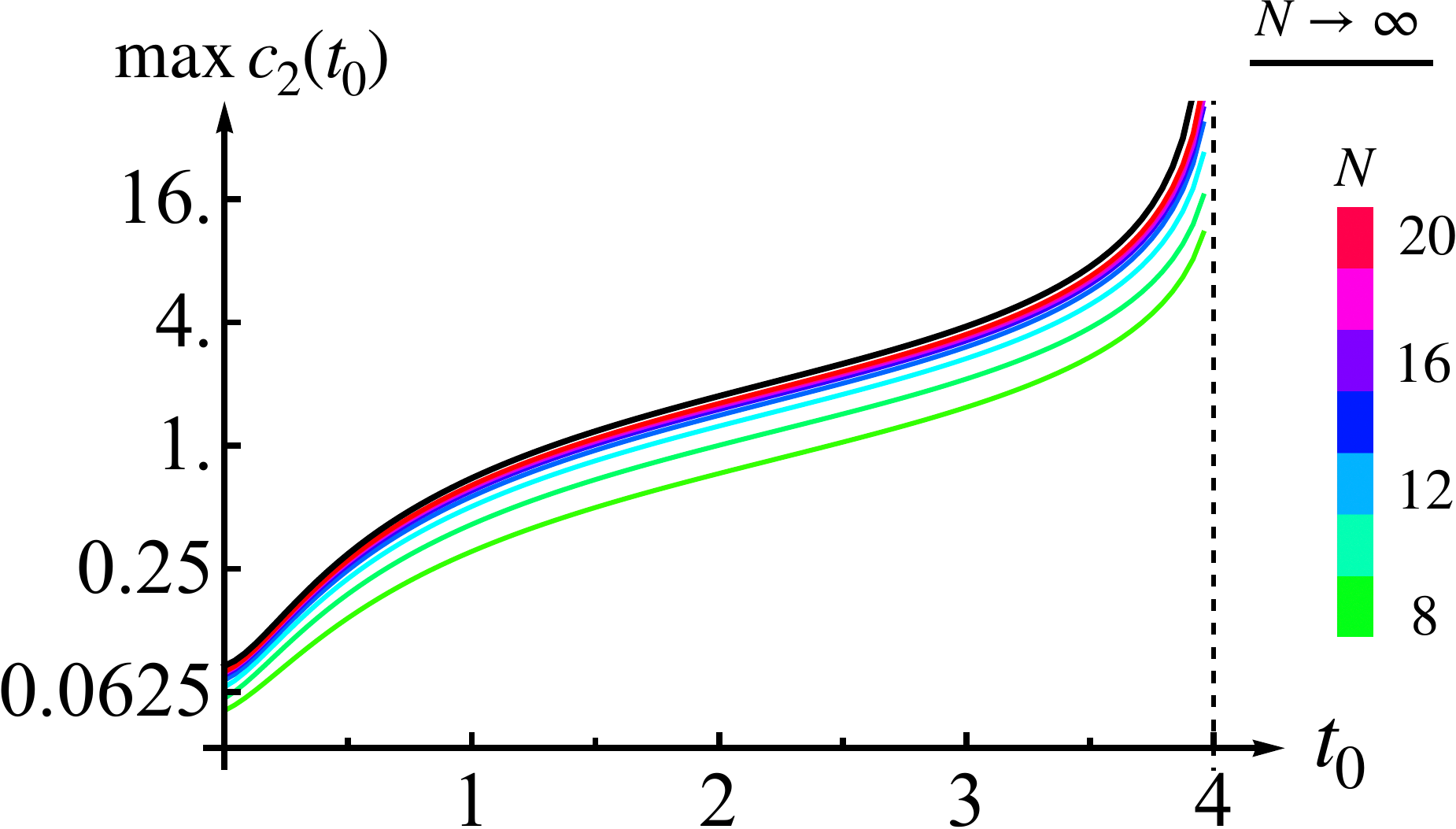}
    \caption{$\text{Max}\, c_2$ as a function of the fixed momentum transfer $0<t_0<4$. Different colors correspond to different values of the cutoff $N$. The extrapolation to $N\to\infty$ is in black. The coefficient $c_2(t_0)$ appears to diverge as $t_0 \to 4$ where it reduces to the spin-2 scattering length \cite{Correia:2020xtr}.}
    \label{fig:max_c2}
\end{figure}

\noindent The upper bound \eqref{eq:sigmatotanalytic} on the integrated cross-section depends on the low energy coefficients $c_{2k}(t_0)$ defined in \eqref{eq:c2_bare}, which can be constrained numerically via the S-matrix Bootstrap.
The obtained maximal value of $c_2(t_0)$ is shown in Figure \ref{fig:max_c2} as function of $t_0$.\footnote{The dual bound on $c_2(4/3)$ was studied in \cite{Gumus:2023xbs,Reggeworkinprogress} using the method developed in \cite{Guerrieri:2021tak}. The bound and our extrapolation differ by 5\%.}

It turns out that for any $t_0$, the amplitude that saturates the maximal value of $c_2(t_0)$ is \emph{universal}. 
This universality extends to all coefficients $\max c_{2k}(t_0)$, suggesting that the amplitude maximizing these coefficients occupies a distinguished position in the space of consistent amplitudes.

\begin{figure}[t]
    \centering
    \includegraphics[width=1.05\linewidth]{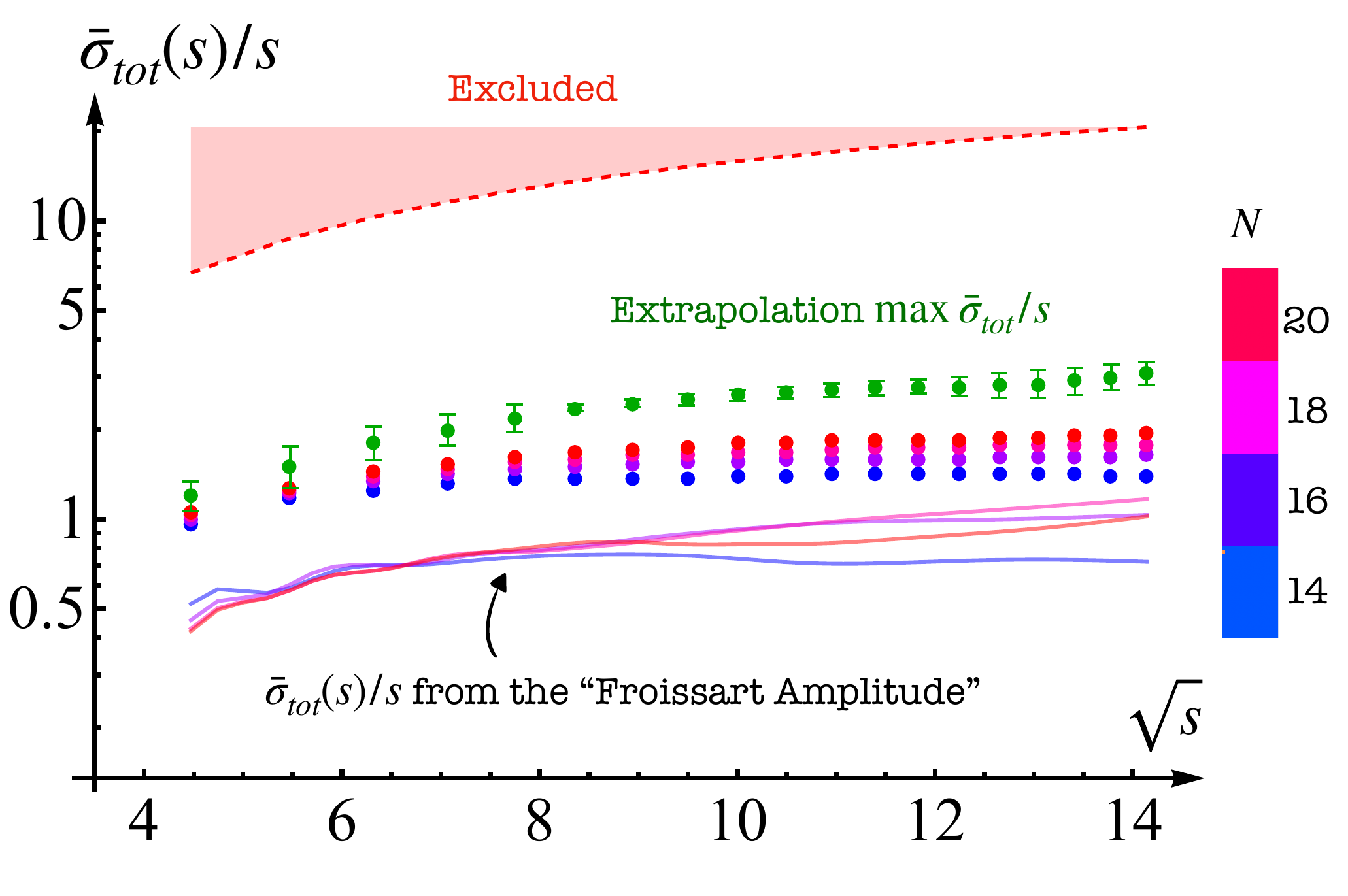}
    \caption{
The dots are obtained by maximizing $\bar\sigma_\text{tot}(s)$ for each value of $s=20,30,\dots,200$.
 Different colors correspond to different values of the ansatz cutoff $N$. The green points with error bars extrapolation to $N\to \infty$ performed with a simple power law fit, which gives an estimate of the optimal numerical bound. In the same figure we show the best analytic bound (red), and the value of $\bar\sigma_\text{tot}(s)$ extracted from the Froissart amplitude.} 
    \label{fig:max_yndurain}
\end{figure}

Having access to $c_{2k}(t_0)$, we can now numerically minimize the objective $\Sigma_4(s,t_0,k,L)$ at fixed $s$.
The best bound is obtained for $k=1$ using the maximal value of the $c_2(t_0)$ coefficient plotted in Figure \ref{fig:max_c2}. The resulting exclusion plot is shown in Fig.~\ref{fig:LHC_vs_us} and \ref{fig:max_yndurain}.

In contrast to the asymptotic Froissart bound~\eqref{eq:Froissart}, where choosing \( t_0 = 4 \) yields the strongest bound, this choice is no longer optimal for the finite energy bound \eqref{eq:sigmatotanalytic} due to the growth of $c_2(t_0)$. In fact, \( c_2(t_0) \) diverges as \( t_0 \to 4 \) as can be seen in Figure \ref{fig:max_c2}. This behavior is intriguing and aligns with numerical evidence indicating that the universal amplitude maximizing these coefficients has a spin-two bound state precisely at threshold. 

\subsection{Direct Maximization of the Cross-Section}
\label{sec:max_cross_section}

\noindent We now maximize the integrated cross-section $\bar \sigma_\text{tot}(s)$ directly with the S-matrix Bootstrap.
This is a  particularly challenging computational problem.\footnote{This can be explained by the fact that the objective $\bar \sigma_\text{tot}(s)$ assumes values on the physical scattering region $s \geq 4$, which is at the boundary of the unit disk where the Taylor series \eqref{eq:foliation} is not guaranteed to converge} Our results are reported in Figure \ref{fig:max_yndurain}. 
The analytic exclusion bound and our extrapolation differ by a factor slightly below five, which is a natural outcome, since the S-matrix Bootstrap amplitude satisfies full crossing symmetry and unitarity at all energies, unlike the analytical derivation.

\begin{figure}[t]
    \centering
    \includegraphics[width=0.9\linewidth]{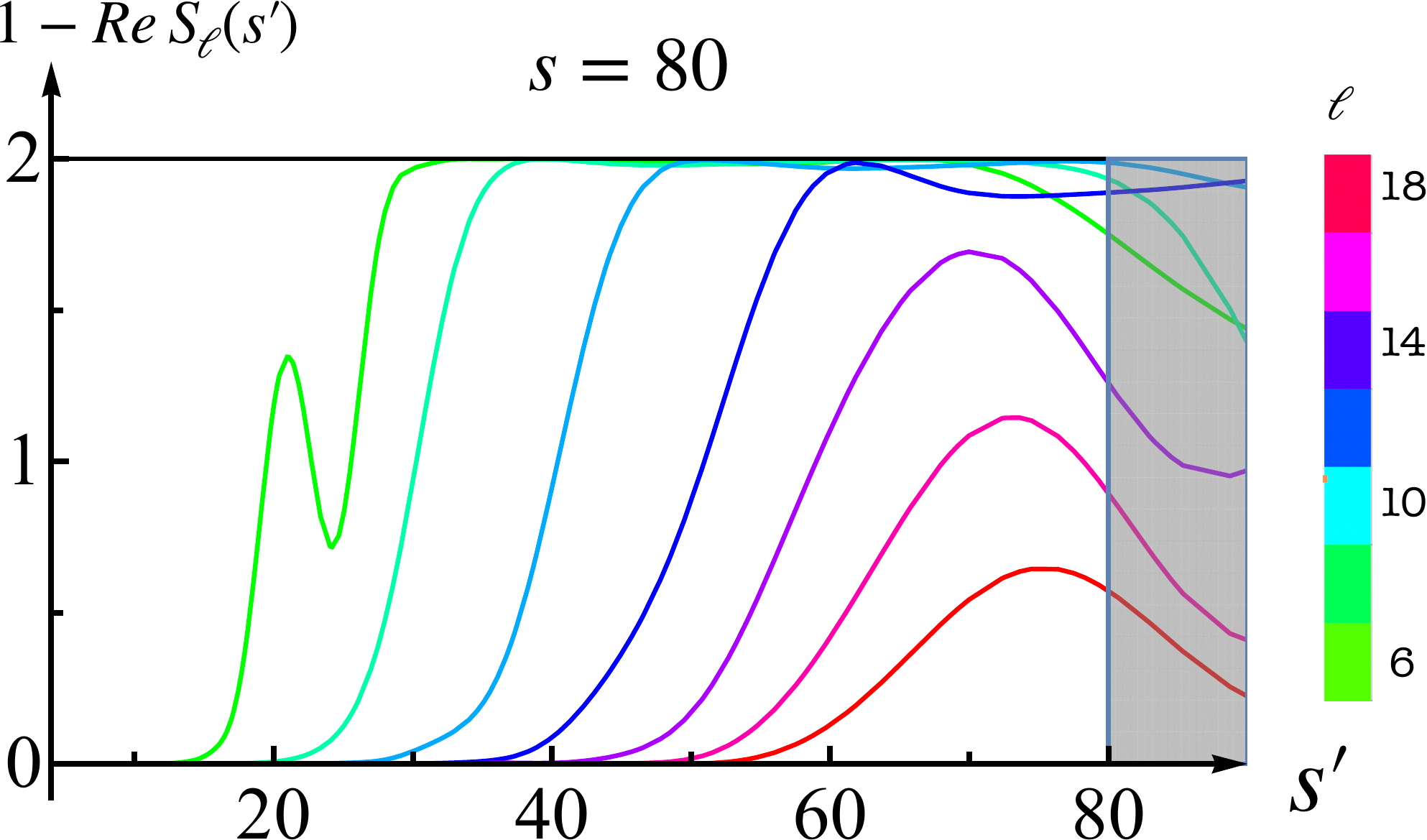}
    \caption{Profile of $1-\Re S_\ell(s^\prime)$ computed from the amplitude maximizing $\bar\sigma_\text{tot}(80)$ for $\ell=6,\dots,18$. This amplitude has resonances arranging into an approximately linear Regge trajectory up to spin $\ell=14$. Above this level, partial waves accumulate close to the boundary of the integration domain resembling a Coon-like behaviour \cite{Coon:1969yw,Figueroa:2022onw}. As we increase $s$, the trajectory extends to larger spins suggesting a regular limit when $s\to \infty$ compatible with the Froissart amplitude proposed in this Letter. The limit however is reached slowly.}
    \label{fig:coon_vibes}
\end{figure}

In Figure \ref{fig:coon_vibes}, we plot the profile of $1-\Re S_\ell(s^\prime)$ for the amplitude maximizing $\bar \sigma_\text{tot}(80)$, focusing on $\ell\geq 6$. 
We find that many of the higher-spin partial waves are appreciably large, therefore violating the commonly observed \emph{low-spin dominance} \cite{Bern:2021ppb}. 
This is expected here given that the saturation of the analytical bound requires $1-\Re S_\ell(s^\prime)=2$ for $\ell < L^*$, where the optimal $L^* = L^*(s)$ increases with energy $s$. In the high-energy limit it can be estimated, where $L^*(s)\simeq {1 \over 4} \sqrt{s} \log s $, leading to the Froissart bound \eqref{eq:Froissart} (see Appendix \ref{sec:Analytical_bound}).\footnote{Understanding how to better model the optimal $L^*$ -- for example extracting it from the numerical partial wave profile, as shown in Figure \ref{fig:coon_vibes} -- could help clarify the origin of the factor-of-five discrepancy between the numerical and analytical bounds.}

We now identify the extremal amplitudes that maximize the integrated total cross-section  $\bar{\sigma}_\text{tot}(s)$ for varying values of $s$, locating them within the allowed region of low-energy coefficients previously explored in \cite{Paper3,He:2021eqn,EliasMiro:2022xaa}.  In those works, the leading coefficients $\{c_0,c_2\}$ were evaluated at the crossing-symmetric point $s = t = u =4/3$, where ${c_0 \over 32 \pi}  \equiv  T(4/3,4/3)$ and $c_2 \equiv c_2(4/3)$ with $c_2(t_0)$ already defined in equation \eqref{eq:c2_bare}.

As shown in Figure~\ref{fig:c0c2}, increasing the energy \(s\) in the maximization of \(\bar{\sigma}_\text{tot}(s)\) causes the low-energy coefficients \((c_0, c_2)\) of the corresponding extremal amplitude to progressively move towards a cusp on the boundary of the allowed region, characterized by simultaneously minimal \(c_0\) and maximal \(c_2\).
The amplitude located at this cusp is the universal amplitude maximizing all low energy coefficients $c_{2k}(t_0)$ discussed in Section \ref{sec:bound_on_integrated}.\footnote{This cusp has been conjectured to correspond to the most strongly-coupled gapped theory in four dimensions \cite{Guerrieri:2021tak, EliasMiro:2022xaa, Gumus:2023xbs}.} 

We conjecture that in the limit $s\to\infty$, the amplitude that maximizes $\bar\sigma_\text{tot}(s)$ coincides with this universal amplitude, which we refer to as the \emph{Froissart amplitude}. In the following section, we investigate the Froissart amplitude in detail and unveil the rich physical structure it encodes.

\section{The Froissart Amplitude}
\label{sec:Froissart_amplitude}

\begin{figure}
    \centering
    \includegraphics[width=\linewidth]{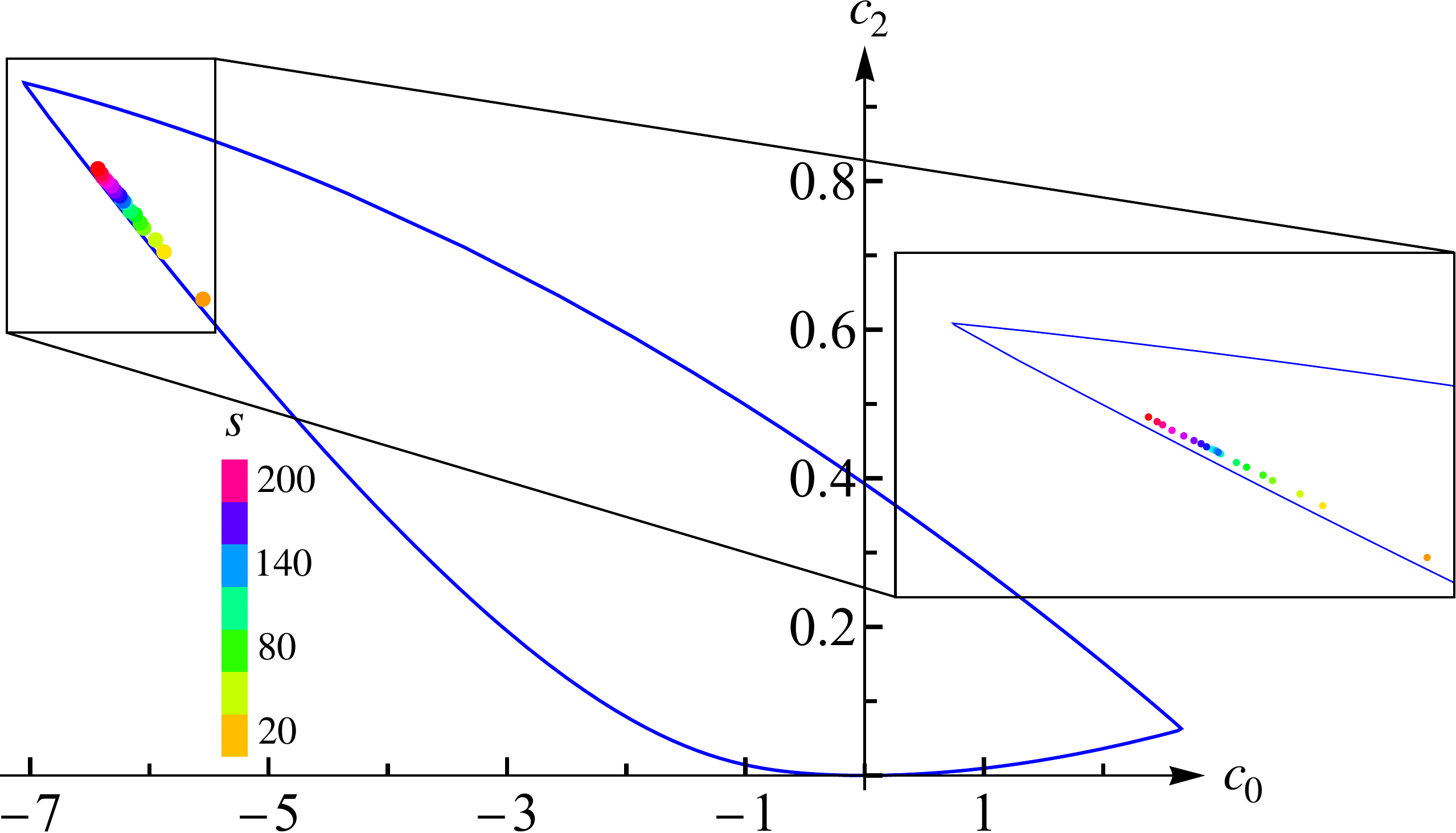}
    \caption{Values of $(c_0, c_2)$ extracted from the amplitudes maximizing $\bar\sigma_\text{tot}(s)$ for $s=20,30,\dots,200$. Different colors correspond to different values of $s$. The blue boundary denote the allowed region, and values outside are not permitted.}
    \label{fig:c0c2}
\end{figure}

\noindent Here we study the Froissart amplitude, which is the extremal amplitude at the $(\min c_0, \max c_2)$ cusp in Figure \ref{fig:c0c2}. 
 We obtain it by directly maximizing the coupling \( c_2 \equiv c_2(\tfrac{4}{3}) \) using the numerical S-matrix Bootstrap described in the previous section.

\subsection{Total Cross-Section}

\noindent The amplitude exhibits a growing cross-section, as shown in Figure~\ref{fig:max_yndurain}, where \(\bar{\sigma}_\text{tot}(s) / s \sim \sigma_\text{tot}(s)\). For each fixed \(s\), the integrated cross-section of this amplitude lies below the result from direct numerical maximization of \(\bar{\sigma}_\text{tot}(s)\) (extrapolated in green) and the analytical upper bound (shown in red). We believe that the integrated cross-section \(\bar{\sigma}_\text{tot}(s)\) of the Froissart amplitude amplitude will asymptotically approach the maximal allowed value as $s \to \infty$.

\begin{figure}[t]
    \centering
    \includegraphics[width=1\linewidth]{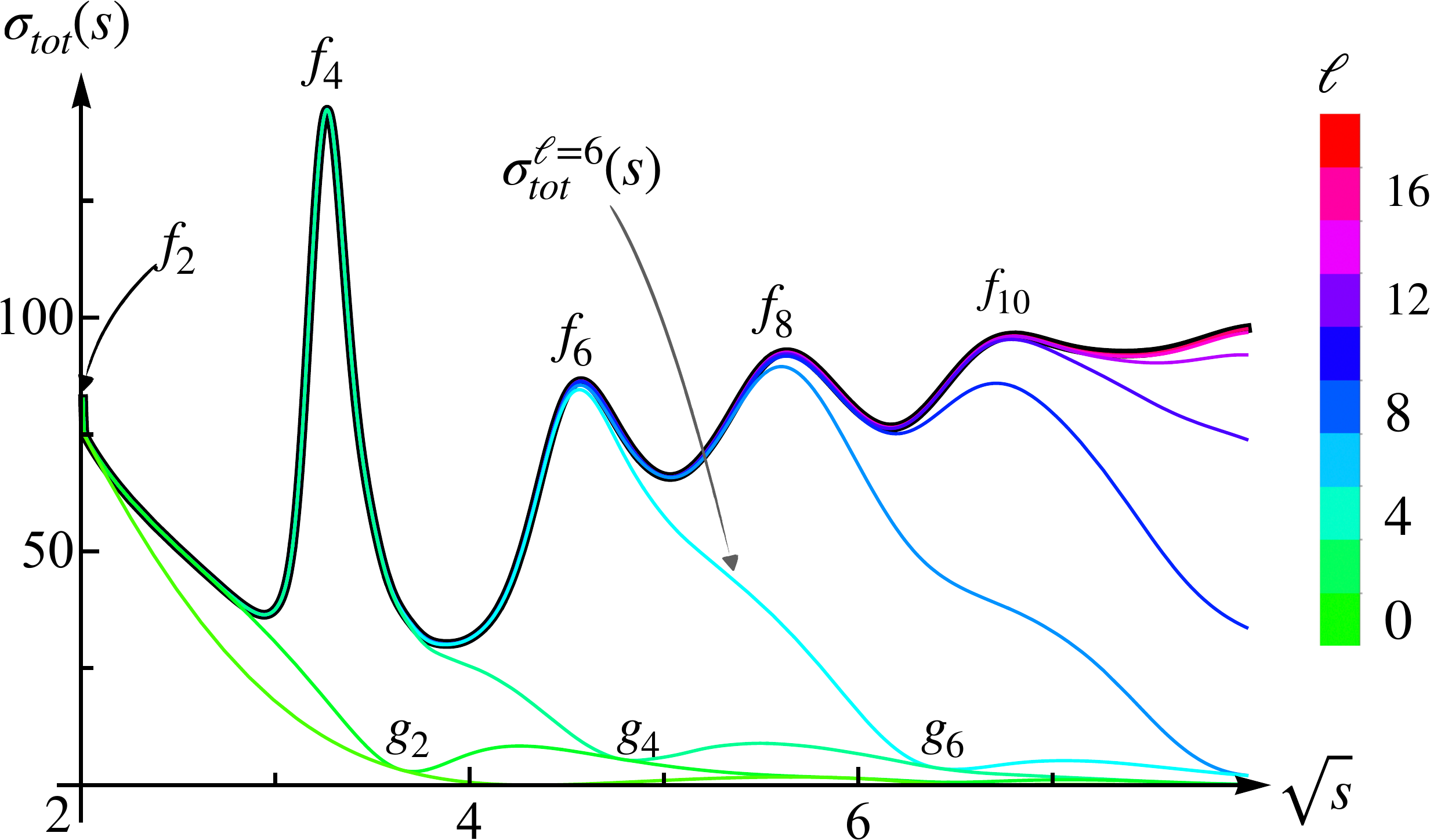}
    \caption{Profile of the total cross-section of the Froissart amplitude for $N=20$. We show the energy window where numerics almost perfectly converged. In different colors we denote the cumulative partial-wave cross-sections, $\sigma_\text{tot}^\ell(s){=} \sum_{\ell'=0}^\ell\frac{16\pi(2\ell'+1)}{s-4}(1{-}\Re S_{\ell'}(s))$, for different spins to highlight the particle content. The peaks, denoted by $f_\ell$, can be correctly identified with resonances particles belonging to Regge trajectory $A$ in the Chew-Frautschi plot in Figure \ref{fig:chew_frauschi}.}
    \label{fig:Froissart_total_sigma}
\end{figure}

A more refined analysis can be carried out by examining the cross-section \(\sigma_\text{tot}(s)\) directly, as shown in Figure~\ref{fig:Froissart_total_sigma}. The plot exhibits a distinct sequence of resonances, denoted \( f_n \), characterized by increasing masses and widths. These resonances collectively drive the overall rise in the cross-section, a trend that is evident in the smeared quantity \( \bar{\sigma}_\text{tot}(s) \).

The partial wave decomposition of the cross-section $\sigma_\text{tot}(s)$, shown in color using equation~\eqref{eq:cross-section}, reveals that each peak is dominated by a single spin contributing almost exclusively at a characteristic energy scale. These observations strongly suggest that the resonances \(f_n\) lie along a Regge trajectory.

Additional features, including secondary bumps and dips, are visible in the partial wave profiles, for instance, the minima denoted by \(g_n\). This ``fine-structure'' of the cross-section hints at a rich spectrum of resonances in the complex plane, which we will now determine directly, leveraging the manifest analyticity of our basis expansion \eqref{eq:foliation}.

\subsection{Chew-Frautschi plot}

\noindent In this section we determine the Chew-Frautschi plot of the Froissart amplitude. To identify the location of a resonance we solve $S_\ell(s)=0$ for complex $s$, corresponding to a pole in the second Riemann sheet.

Owing to the analyticity of our expression \eqref{eq:foliation} we can further analytically continue the partial wave away from integer $\ell$ via the Froissart-Gribov formula (see e.g. \cite{Acanfora:2023axz} for details),
\be
\label{eq:FG}
S_\ell(s)=1\!+\!\frac{i}{4 \pi^2}\int_4^\infty Q_\ell(1+\tfrac{2t}{s-4}) \, \frac{\Disc_t T(s,t)} {\sqrt{s(s-4)}}\,dt.
\ee

\begin{figure*}
    \centering
    \includegraphics[width=\linewidth]{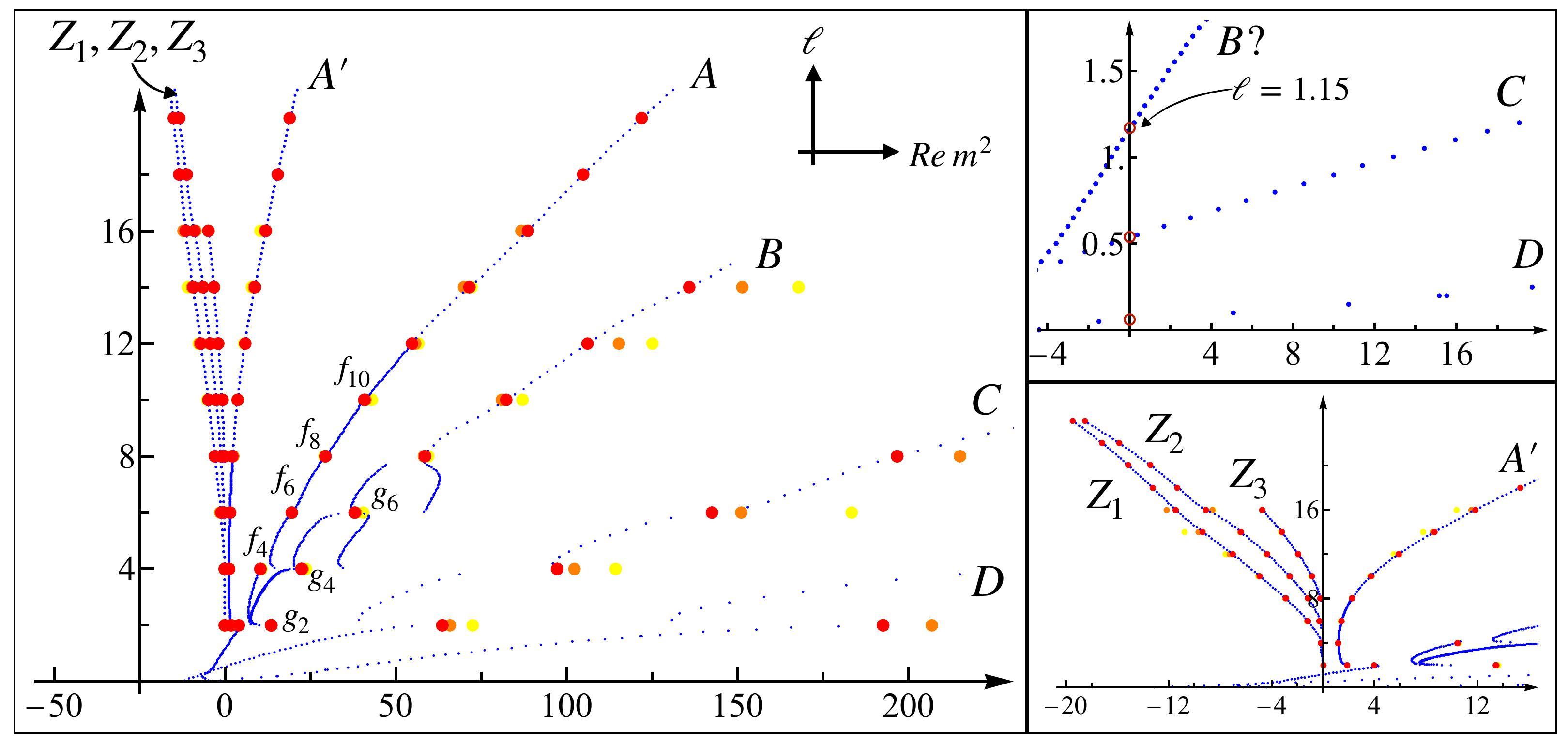}
    \caption{Chew-Frautschi plot of the Froissart amplitude. Dots colored in red, orange, and yellow denote even integer spins respectively for $N=20,18,16$, corresponding to physical particles. Blue dots denote the particles showing up at real values of the spin arranging into Regge trajectories. We denote with $f_\ell$, and $g_\ell$ the visible in Figure \ref{fig:Froissart_total_sigma}, and detectable from the profile of the cross-section.
    The rest of the particles are undetectable looking at the physical region -- this raises the philosophical question \emph{what is a particle?}
    The trajectories $Z_i$, and $A^{(\prime)}$, are reminiscent of trajectories predicted in the 1970s in connection to the saturation of the Froissart bound as discussed in the text.  These are stable in $N$, while the remaining trajectories appearing at large $s$ are still far from converging. The cusps and branchings of the analytically continued spin are likely due to convergence issues. The inset at the top-right corner, shows the three trajectories intersecting the real axis $\Re m^2=0$. We are tempted to identify $B$ as the \emph{Pomeron} trajectory, due to its intercept close to one $\ell\sim 1.15$. Sub-leading intercepts are at $\ell\sim0.53$, and $\ell\sim 0.06$. }
    \label{fig:chew_frauschi}
\end{figure*}

\noindent The result of our analysis is shown in the Chew-Frautschi plot in Figure \ref{fig:chew_frauschi}. We observe a variety of resonances which align along approximately smooth Regge trajectories, up to the expected numerical uncertainty. Among these, trajectory $A$ stands out immediately: it contains the resonances $f_n$ responsible for the pronounced peaks in the total cross-section shown in Fig.~\ref{fig:Froissart_total_sigma}. At large $s$, this trajectory is well-described by the asymptotic form $\ell \sim \sqrt{s} \log s$, which is consistent with an amplitude saturating the Froissart bound \eqref{eq:Froissart} (see Appendix \ref{sec:Analytical_bound}).

The rightmost trajectories---labeled $B$, $C$, and $D$---converge more slowly. Some of the faint ``dips" in the cross-section plot, denoted $g_n$, can be attributed to resonances on  trajectory $B$.\footnote{A similar phenomenon was recently observed in $\pi\pi$ scattering~\cite{Guerrieri:2024jkn}.} 

As shown in the upper-right panel of Fig.~\ref{fig:chew_frauschi}, we find that trajectories $B$, $C$, and $D$ intersect the $\Re m^2 = 0$ axis. According to Regge theory (see Appendix \ref{sec:regge} for a brief introduction), the highest intercept $\alpha_0$ governs the high-energy behavior of the forward scattering amplitude, $T(s, 0) \sim s^{\alpha_0}$, and consequently, the total cross-section $\sigma_{\text{tot}}(s) \sim s^{\alpha_0 - 1}$. A rising cross-section thus requires $\alpha_0 \geq 1$, with the corresponding trajectory traditionally referred to as the \emph{Pomeron} in QCD~\cite{Chew:1962eu,Gribov:1968uy}.

From our Chew-Frautschi plot, the  Pomeron could be associated with trajectory $B$ as it is well-fitted by a linear trajectory\footnote{From here on we will identify $\Re m^2$ with $t$ when talking about resonances as usually done in Regge theory.} $\alpha(t)=\alpha_0+\alpha^\prime t$, with $\alpha_0\simeq 1.1$, and $\alpha^\prime\simeq 0.2$. It is curious that these values are numerically close to the experimental values extracted from hadronic collisions \cite{Covolan:1996uy, Luna:2003kw}.\footnote{Coincidentally, they are also close to the values extracted in \cite{Acanfora:2023axz}.} An independent fit of the growth of our amplitude yields a similar value for $\alpha_0 \simeq 1.2$ (see Appendix \ref{subsec:growatinf}).\footnote{Note that compliance with the Froissart bound \eqref{eq:Froissart} strictly requires $\alpha_0 \leq 1$, and similarly to experimental fits in  proton-proton scattering, the values extracted for $\alpha_0$ are only believed to hold for a transient energy range.}

We also identify several unconventional trajectories, labeled \( A' \), \( Z_1 \), \( Z_2 \), and \( Z_3 \).\footnote{Similar trajectories were first observed in the context of supergravity amplitudes~\cite{Guerrieri:2022sod} and further investigated in~\cite{Guerrieri:2023qbg}.} Among these, the \( A' \) trajectory is particularly notable for exhibiting a turning point, as shown in the bottom-right inset. Trajectories with similar behavior in the forward limit—specifically those of the square-root form \( \alpha(t) = 1 + r_0 \sqrt{t} \)— were first proposed by J. Schwarz \cite{Schwarz:1968zz} and have subsequently appeared in few other instances within the classical Regge theory literature~\cite{Sawada,OEHME1970573,Gribov:1969ig,Bronzan:1971cyt,Bronzan:1972wd} (see also Sec.~8.3 of~\cite{Collins:1977jy}), particularly in connection with the saturation of the Froissart bound~\cite{Oehme:1971omw,Oehme:1973ch,Pakvasa:1974av,Dubovikov:1976ub, Kupsch:1982aa,Kupsch:2008hq}.

Indeed, as we will now show, a simple analytical model featuring such a trajectory captures several features of the Froissart amplitude obtained from the numerical S-matrix Bootstrap.

\subsection{Elastic Differential Cross-Section}

\begin{figure}
    \centering
\includegraphics[width=\linewidth]{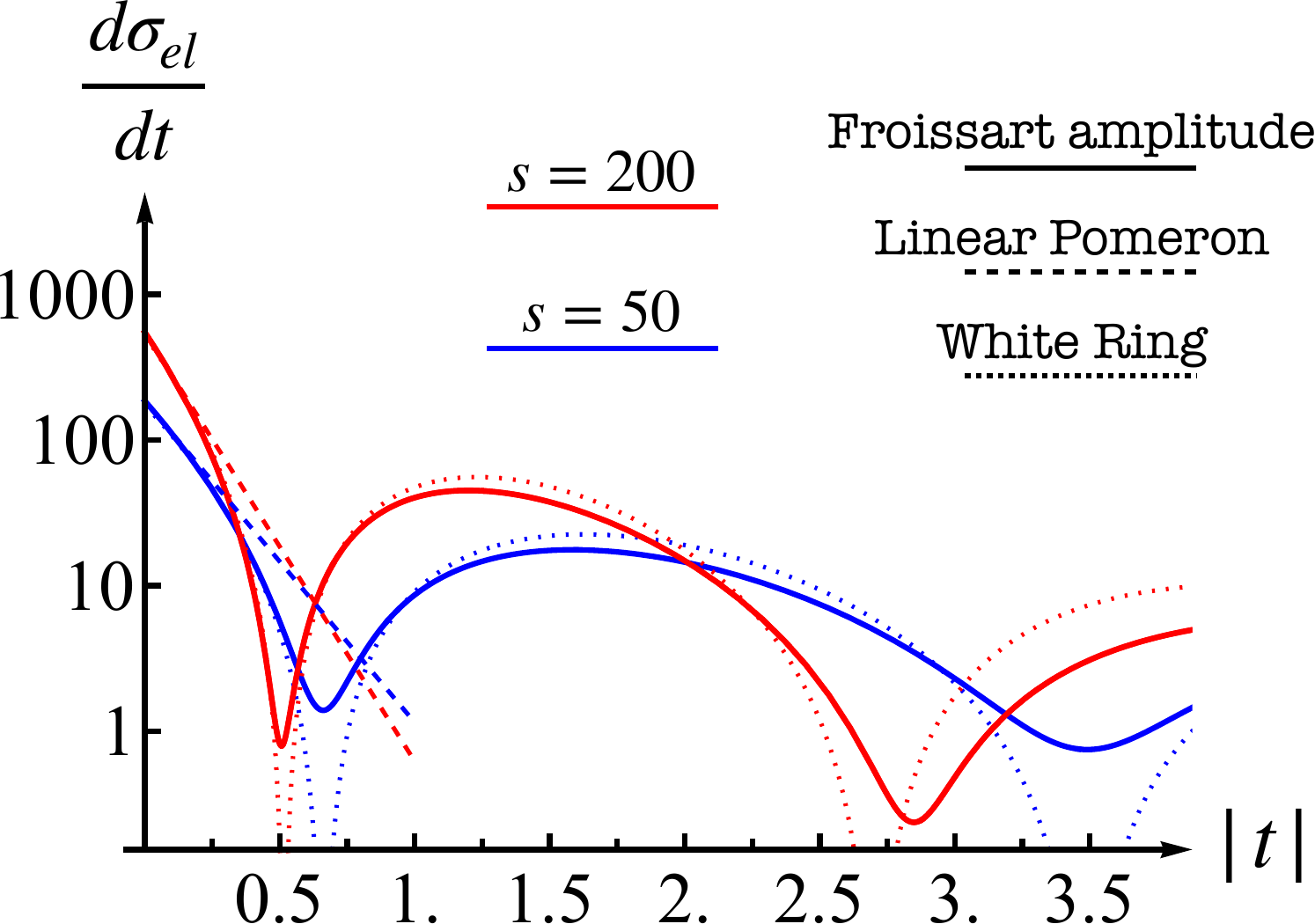}
    \caption{Diffractive cone of the Froissart amplitude. It exhibits exponential fall-off at small $|t|$ consistent with  a linear Pomeron trajectory (dashed), and at larger $|t|$ quasi-periodic ``dips" reminiscent of a black disk diffraction pattern (dotted). The cone shrinks with increasing energy $s$ as can be seen from the two chosen values $s = 50 , 200$.}
    \label{fig:diffractive_cone}
\end{figure}

\noindent In this section, we study another important observable in high-energy hadronic collisions: the elastic differential cross-section,
\be
\label{eq:elasticsigma}
{d \sigma_{el} \over dt} = {|T(s,t)|^2 \over 16 \pi s ( s - 4)}.
\ee
 We plot it in Figure \ref{fig:diffractive_cone} for physical $t \leq 0$ and two values of energy $s = 50$ (blue) and $s = 200$ (red).

We observe that the elastic differential cross-section grows rapidly with $s$ in the forward limit \( t \to 0 \), followed by a steep decline for \( t \lesssim 0 \). As the energy \( s \) increases, this fall-off in \( |t| \) becomes even more pronounced. This characteristic behavior has been well established in proton-proton scattering since the 1970s~\cite{Collins:1977jy, Gross:2022hyw, ParticleDataGroup:2024cfk}, and is commonly referred to as the ``shrinking of the diffractive cone''.

Explaining this behavior is regarded as one of the major successes of Regge theory~\cite{Collins:1977jy,Donnachie_Dosch_Landshoff_Nachtmann_2002}. If at high energies the amplitude is dominated by a single linear Pomeron trajectory, \( T(s,t) \sim s^{\alpha_0 + \alpha' t} \) with \( \alpha' > 0 \), the elastic differential cross-section behaves as
\be
\label{eq:linpom}
\frac{d \sigma_\text{el}}{dt} \propto s^{2(\alpha_0 - 1)} e^{-(2 \alpha' \log s) |t|}.
\ee
The {slope parameter}, given by $2 \alpha' \log s$, increases slowly with energy, thereby producing the shrinking of the cone. In Figure \ref{fig:diffractive_cone} we observe a similar behavior for small $|t| \lesssim 0.25$, where an exponential fit is superimposed (shown as a dashed line), even though the shrinking rate appears faster than $\log s$.

For values of \( |t| \sim 0.5 \), we observe the emergence of a minimum, which shifts to smaller \( |t| \) and becomes more pronounced as the energy increases. 
In Figure~\ref{fig:diffractive_cone}, we show the appearance of a second minimum which, like the first, shifts toward smaller \( |t| \) and becomes sharper with increasing energy.

\begin{figure}
    \centering
\includegraphics[width=\linewidth]{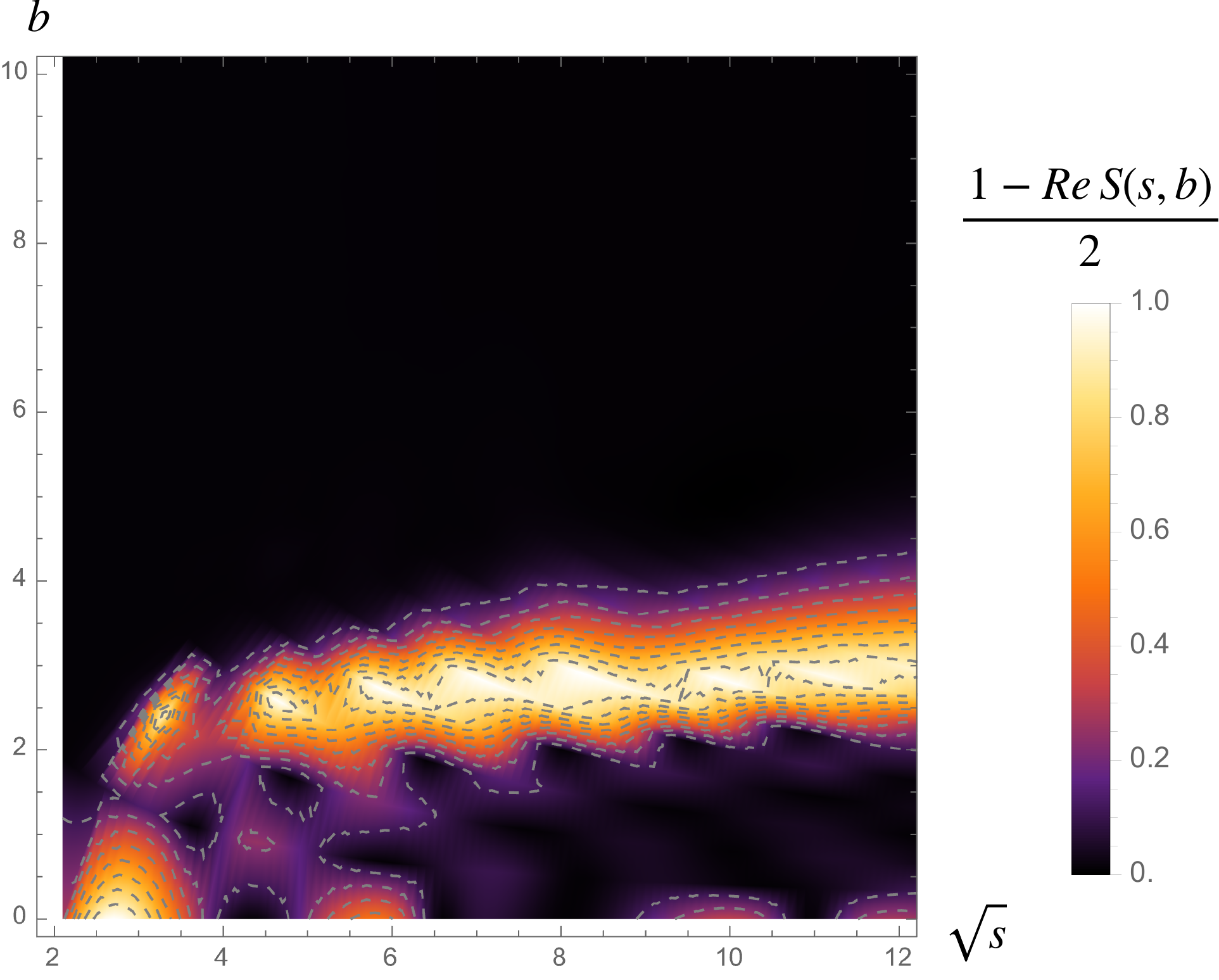}
    \caption{Density plot of the eikonal profile of the Froissart amplitude. Most of the scattering takes place within a finite interval in impact parameter resembling a ring profile. The rising cross-section is a consequence of the slow expansion of the ring with energy. This is apparent by fitting the ``white ring" model in \eqref{eq:whitering} to the diffractive cone profile in Figure \ref{fig:diffractive_cone}. We find that inner radius is essentially unchanged $R_1(s=50)=R_1(s=200) \simeq 2.3$ while outer radius grows slowly $R_2(s=50) \simeq 3.5$ and $R_2(s=200) \simeq 4.4$, compatible with Froissart growth.}
    \label{fig:whitering}
\end{figure}

Physically, these oscillations resemble the Fraunhofer pattern produced by a beam of light passing through an annular aperture, known as an obscured Airy pattern~\cite{born1980principles,2017opti}. In Figure~\ref{fig:whitering}, a Fourier transform of the Froissart amplitude indeed confirms the emergence of a ring-like structure, where the exchanged momentum \( \sqrt{-t} \) is conjugate to the impact parameter \( b \).

Using the eikonal representation (see Appendix~\ref{sec:whitering}), we model this ring structure by assuming perfect reflection for impact parameters in the range \( R_1 < b < R_2 \), and no interaction elsewhere, where \( R_1 \) and \( R_2 \) denote the inner and outer radii, respectively. This leads to the following ``white ring'' diffraction model:
\be
\label{eq:whiteringd4}
T_\text{WR}(s,t) = \frac{8 \pi i s}{\sqrt{-t}} \left[ R_2 J_1(R_2 \sqrt{-t}) - R_1 J_1(R_1 \sqrt{-t}) \right],
\ee
where \( J_1 \) is the Bessel function of the first kind.

In Figure~\ref{fig:diffractive_cone} (dotted lines), we show the elastic differential cross-section~\eqref{eq:elasticsigma} predicted by the white ring model~\eqref{eq:whiteringd4}, with the parameters \( R_1 \) and \( R_2 \) fitted to better match the Froissart amplitude. We find that the white ring model offers an excellent description of the diffractive cone’s fall-off at small \( |t| \), outperforming the linear Pomeron fit~\eqref{eq:linpom}. It also captures reasonably well the locations of the minima in the Froissart amplitude, where it instead has zeros.\footnote{The presence of minima rather than zeros in the Froissart amplitude is likely due to the contribution of subleading trajectories with a softer fall-off in \( |t| \), as seen in the top-right inset of Figure \ref{fig:chew_frauschi}. These ``background" contributions become increasingly suppressed at higher energies due to the \( 1/s^2 \) factor in~\eqref{eq:elasticsigma}, leading to deeper minima, as seen in Figure~\ref{fig:diffractive_cone}.}

The fitted values of \( R_1 \) and \( R_2 \) are captioned in  Figure~\ref{fig:whitering}, and match roughly with the eikonal profile of the Froissart amplitude. We observe that \( R_2 = R_2(s) \) increases slowly with energy, consistent with a growing total cross-section. Note that the expansion of the ring causes shrinking of the \emph{entire} diffraction pattern, as seen in Figure~\ref{fig:diffractive_cone}.

The total cross-section of the white ring model can be computed via the optical theorem \eqref{eq:optical}
\be
\label{eq:sigmaring}
\sigma_\text{tot,WR}(s) = 4 \pi \big[(R_2)^2 - (R_1)^2 \big].
\ee
In this particular model, Froissart growth can be achieved if the outer radius $R_2$ grows logarithmically $R_2(s) = r_0 \log s$. The Froissart bound \eqref{eq:Froissart}, and particularly the upper bound on the spin $L \simeq \sqrt{s \over 4 t_0} \log s$ (see Appendix \ref{sec:Analytical_bound}) place a bound on the rate of growth, $r_0 \leq 1/\sqrt{t_0}$, where $t_0$ is the closest singularity in the $t$-channel, which is $t_0 = 4$ in our case.

If the inner radius $R_1$ also grows logarithmically the ring structure would be preserved at high energies. This would imply that the overall coefficient of $\sigma_\text{tot}(s) \propto \log^2 s$ has to be smaller than the factor $4 \pi /t_0$ of  the Froissart bound \eqref{eq:Froissart}. It would be interesting to see if this is the case by going to higher energies with the numerical S-matrix Bootstrap. 

Interestingly, we can also examine the white ring model~\eqref{eq:whiteringd4} in the crossed channel. Assuming 
the Froissart behavior \( R_2 = r_0 \log s \), we can determine the leading singular behavior of the $t$-channel partial wave amplitude \( S_\ell(t) \) in the region \( \ell \sim 1 \) and \( t \sim 0 \) (see Appendix~\ref{sec:whitering} for details). We find
\be
\label{eq:pwt}
S^\text{WR}_\ell(t) \sim \left[ (\ell - 1)^2 - r_0^2 t \right]^{-\frac{3}{2}}.
\ee
The Regge trajectory of the white ring model is therefore a Regge cut, with branch points at \( \ell = 1 \pm r_0 \sqrt{t} \), which coalesce into a third-order Regge pole at the intercept \( \ell = 1 \), as expected from Froissart growth (see Appendix~\ref{sec:regge}). This steep behavior of the trajectory near \( t \to 0 \) resembles what we observe in the Chew–Frautschi plot (bottom-right inset of Figure~\ref{fig:chew_frauschi}), although the two pictures do not align exactly. This discrepancy could well be due to numerical limitations,\footnote{ 
We believe that the analytic continuation of partial waves to $\ell \sim 1$ may still be affected by systematic errors due to numerical convergence, as it is also sensitive to the behaviour at large $s$. 
} 
but it is also worth noting that the simple analytical model~\eqref{eq:pwt} cannot hold exactly: its very singular behavior in \( t \) would violate analyticity and unitarity in the \( t \)-channel. It is possible that the  Froissart amplitude has a much more complicated mechanism, potentially involving ``hiding cuts"~\cite{Oehme:1973ch} which could be related to the presence of the multiple $Z_i$ trajectories we see in Figure \ref{fig:chew_frauschi}. We leave this fascinating question to the future.

In Appendix \ref{appendix:Froissart_amplitude} we collect additional data of the Froissart amplitude in the complex plane: the list of the first seven Wilson coefficients, the profile of the phase shifts, and the existence of complex zeroes of the amplitude close to the negative $t$-axis which are responsible by the diffractive minima and whose presence was predicted long-ago \cite{Auberson:1971ru}  as a universal consequence of cross-section growth.

\section{Discussion and outlook}
\label{sec:conclusion}

In this work, we have derived analytic and numerical bounds on the integrated total cross-section for the scattering of scalar particles, valid at all energies and in arbitrary spacetime dimensions. These bounds rely solely on general S-matrix principles: causality, crossing symmetry, and unitarity. At high energies, where spin effects become negligible, our universal analytic bound lies within a factor of ten of experimental $pp$ and $p \bar{p}$ scattering data, when setting $m = 1 \, \text{GeV}$. 

We propose that the Froissart amplitude identified in this work is the function that maximizes the integrated cross-section across all energy scales. This amplitude exhibits a rising total cross-section and an elastic differential cross-section characterized by a diffractive pattern with multiple minima that shrinks with increasing energy. These features arise from a complex spectrum of resonances lying on Regge trajectories, which notably includes a spin-2 bound state at the two-particle threshold.

Remarkably, the bootstrap indicates that a rising cross-section  is not achieved via a growing disk, as traditionally expected, but rather through a \emph{growing annulus}. This observation is supported both by the eikonal profile of the Froissart amplitude and the observed diffractive pattern in the differential cross-section.

The Regge spectrum of the Froissart amplitude, determined directly through the manifest analyticity of the bootstrap construction, also reflects this picture. The Chew-Frautschi plot features linear trajectories at small \( t \) that bend at higher values, consistent with Froissart-like behavior $\sim \sqrt{t} \log t$. Interestingly, moreover, in the forward region \( t \sim 0 \), we observe trajectories with ``singular" behavior, including one with an \emph{turning point}, echoing analytical predictions from the 1970s associated with the saturation of the Froissart bound \cite{Oehme:1971omw,Oehme:1973ch,Pakvasa:1974av,Dubovikov:1976ub,Kupsch:1982aa,Kupsch:2008hq}.

Our analysis, grounded in first principles, suggests that these properties are deeply interconnected and intrinsic to Froissart growth. That Froissart growth provides a good fit of the high-energy $pp$ cross-section~\cite{ParticleDataGroup:2024cfk} raises a tantalizing possibility: could some of the features observed in the Froissart amplitude, also be present in QCD, and be a result of consistency with fundamental principles?

Some similarities could suggest a connection:
\begin{enumerate}
    \item The observed shrinking of the $pp$ diffractive cone at a rate faster than $\log s$ at $\sqrt{s} = 13 \text{ TeV}$ as reported by the TOTEM collaboration \cite{TOTEM:2017asr, ParticleDataGroup:2024cfk}. This is a feature of our Froissart amplitude and of the analytical ``white ring" model with explicit Froissart growth.
    \item  A diffractive-like pattern with multiple minima that becomes more pronounced at higher energies. While only one minimum is currently observed at the LHC, current fits, which use a combination of Regge-pole and eikonal models, typically predict further structures beyond current experimental resolution \cite{Fiore:2008tp, Martynov:2018sga, Jenkovszky:2018itd, Ferreira:2020nyi}. It would be intriguing if, at higher energies, the diffractive cone structure of $pp$ scattering were to more closely resemble the predictions emerging from our fundamental bootstrap approach.
    \item The presence of a spin-2 resonance near threshold in our Froissart amplitude, as seen in Fig. \ref{fig:Froissart_total_sigma}. If this particle belongs to a Regge trajectory that is associated with universal growth, it should have vacuum quantum numbers. Curiously the  $f_2(1950)$ resonance \cite{ParticleDataGroup:2024cfk} matches this description, which has been alternatively proposed as the first state on the Donnachie-Landshoff Pomeron trajectory \cite{Donnachie:1985iz, Donnachie_Dosch_Landshoff_Nachtmann_2002}.

\end{enumerate}

Notably, the Froissart amplitude we construct is predominantly elastic within numerical precision. This is expected: any inelasticity would reduce the total cross-section, given unitarity. In contrast, real-world $pp$ scattering is mostly inelastic, though it has been suggested that the black disk limit, where $\sigma_{\text{el}} / \sigma_{\text{tot}} \sim 0.5$ may be asymptotically approached.

\subsection*{Outlook and future directions}

This work demonstrates for the first time that the S-matrix Bootstrap, traditionally applied to bound low-energy observables, can also be used to explore high-energy behavior.\footnote{See also \cite{Bhat:2023puy} for a previous interesting exploration.}
In particular, we identified the Froissart amplitude, which exhibits all the nontrivial features discussed above, as a consequence of consistency with fundamental principles alone. To our knowledge, this is the most intricate scattering amplitude constructed within the S-matrix Bootstrap framework.

Our results also highlight the potential of combining numerical bootstrap techniques with analytical methods. As shown in this work, such synergy enabled the derivation of a universal bound on a directly measurable quantity: the total cross-section at finite energies. The universality of this approach means it can be applied broadly in different contexts. We expect it to be easily generalizable for  different dimensions, or spins and internal quantum numbers of the scattering particles. Similar arguments could be done for other physical observables.

An obvious next step is to extend the analysis to fermionic amplitudes, particularly to proton-proton scattering. S-matrix Bootstrap studies with fermions were first obtained in \cite{Hebbar:2020ukp}. The study of non-identical fermions could reveal new structures in the Regge spectrum, including the emergence of the  Odderon trajectory \cite{Lukaszuk:1973nt, Donnachie_Dosch_Landshoff_Nachtmann_2002, Martynov:2017zjz,  TOTEM:2018psk}. Bootstrapping a four-fermion amplitude, or a mixed system like pion-proton scattering,\footnote{It has been suggested \cite{ANSELM1972487, Jenkovszky:2017efs} that the $\pi \pi$ threshold could be responsible for the non-exponential rise in the diffractive cone at very small $t$ \cite{TOTEM:2015oop}, called the ``break" \cite{Fiore:2008tp, Jenkovszky:2017efs,Jenkovszky:2018itd}. A dedicated S-matrix Bootstrap study could provide a definitive answer to this longstanding question.}  would allow for a more direct comparison with experimental data, even at low energies. The resulting fermionic Froissart amplitude could potentially provide a closer theoretical proxy for real-world proton scattering.

Another natural direction is to incorporate inelasticity, which arises organically within the bootstrap framework of \cite{Tourkine:2023xtu, Gumus:upcoming}. This prompts a compelling question: rather than maximizing the total cross-section, can one instead \emph{maximize the inelastic cross-section}? Doing so could elevate the black disk model from a mere heuristic expectation to a robust theoretical bound, potentially offering insights into the asymptotic elastic-to-total cross-section ratio.\footnote{Curiously, there are also indications that inelastic scattering becomes increasingly peripheral, a phenomenon referred to as ``hollowness" \cite{RuizArriola:2016tre, Mantysaari:2020axf, Broniowski:2018xbg} or the ``black ring" \cite{Troshin:2022kym} effect. It would be fascinating if this feature could also emerge from consistency conditions alone.}

From a phenomenological perspective, one can introduce additional assumptions to better align bootstrap amplitudes with experimental data. As shown in recent work~\cite{Guerrieri:2024jkn}, the S-matrix Bootstrap provides a fully self-consistent framework for such fits. Applying it to total and elastic differential hadronic cross-sections would yield amplitudes that automatically satisfy analyticity, unitarity, and crossing symmetry. These amplitudes could not only reproduce existing data but also predict the evolution of cross-sections and diffractive patterns at higher energies, including the emergence of new diffractive minima. More importantly, they would allow direct access to the QCD resonance spectrum and its Regge trajectories, including the long-conjectured Pomeron and the hypothetical ``Froissaron''. This has long been the aim of Regge-based models, which typically incorporate only a limited subset of fundamental constraints. Our results suggest that the modern bootstrap offers a path toward a systematic theoretical and phenomenological understanding of \emph{soft} QCD, a regime where few alternative tools exist.

These prospects also resonate with earlier analytical constructions of amplitudes exhibiting Froissart growth. In particular, the simple analytic ring model introduced here shares key qualitative features with a class of amplitudes studied by Kupsch~\cite{Kupsch:1982aa, Kupsch:2008hq} and others in the 1970s~\cite{Oehme:1971omw,Oehme:1973ch,Pakvasa:1974av,Auberson:1971ru}. While these early works were largely theoretical explorations, recent indications of Froissart-like growth in hadronic cross-sections~\cite{ParticleDataGroup:2024cfk, TOTEM:2017asr, COMPETE:2002jcr} invite renewed interest in such analytic approaches and their further development. All current evidence suggests the ``Froissaron'' is a highly nontrivial structure in the S-matrix—likely a composite object involving one or more Regge cuts—pointing to yet another remarkable facet of the nonperturbative dynamics of QCD.

\section*{Acknowledgments}
\noindent We thank Simon Caron-Huot, Simon Ekhammar, Mathieu Giroux, Giulia Isabella, Kelian Häring, Aditya Hebbar, Brian McPeak, João Penedones, Junsei Tokuda, Balt van Rees, Pedro Vieira, and Sasha Zhiboedov for useful discussions and comments on the draft. We are especially grateful to Simon Caron-Huot and Sasha Zhiboedov for their particularly valuable input. M.~C. is supported by the National Science and Engineering Council of Canada (NSERC) and the Canada Research Chair program, reference number CRC-2022-00421. A.~G.~is supported by a Royal Society funding, URF\textbackslash{R}\textbackslash221015. A.~L.~G is supported by a Royal Society University Research Fellowship,  URF\textbackslash{R1}\textbackslash241371.

\begin{appendix}

\section{Point-wise infinite cross-section}
\label{sec:divcross}

\begin{figure}[t!]
    \centering
    \includegraphics[width=\linewidth]{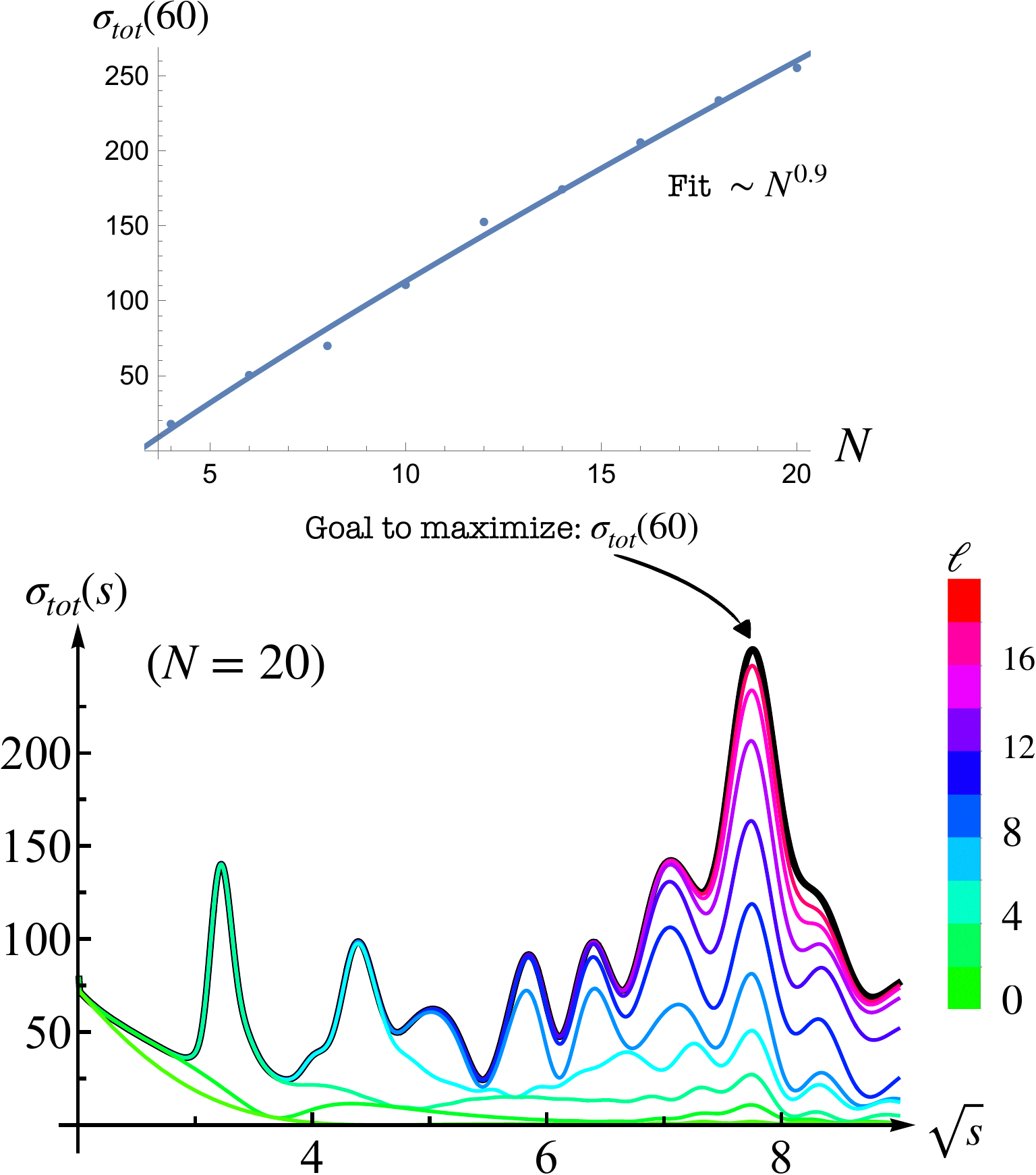}
    \caption{Direct pointwise maximization of the total cross-section at \(s = 60\). As shown in the upper panel, no strict upper bound appears to exist. The lower panel displays the cross-section profile of the extremal amplitude for \(N = 20\). We observe the emergence of a tower of higher-spin resonances clustering around \(s = 60\). These resonances are expected to Reggeize, giving rise to additional resonances at energies away from \(s \sim 60\), as seen in the plot. Interestingly, a universal pattern emerges at lower energies, where similar peaks are found in the Froissart amplitude.}
    \label{fig:max_pointwise}
\end{figure}

In this Appendix, we explicit construct a model that makes the cross-section $\sigma_\text{tot}(s)$ diverge at a single point $s \to s_0$. Consider the following Breit-Wigner model corresponding to a series of higher-spin narrow resonances with the same mass $\sqrt{s_0}$ and width $\Gamma$,
\be
\label{eq:breitwigner}
S_\ell(s) = {s - s_0 - i \Gamma \over s - s_0 + i \Gamma}.
\ee
Near $s \sim s_0$ each will contribute to the cross-section \eqref{eq:cross-section} with
\be
1 - \Re S_\ell(s) = {2 \, \Gamma^2 \over (s - s_0)^2 + \Gamma^2}
\ee
and, in particular, when $s = s_0$ we have saturation of unitarity $1 - \Re S_\ell(s_0) = 2$ for all $\ell$ meaning that the cross-section diverges $\sigma_\text{tot}(s \sim s_0) \propto \sum_{\ell=0}^\infty n_\ell^{(d)} \to \infty$. 

This is a very rough model, and it is likely that for wide enough resonances $\Gamma \sim O(1)$ it violates the upper bound on $c_2$ or on the integrated cross-section $\bar{\sigma}_\text{tot}(s > s_0)$, as described in section \ref{sec:anbound}.

We can refine the model such that $\sigma_\text{tot}(s \sim s_0) \propto \delta(s - s_0)$. This can be achieved by letting each resonance have a very small width $\Gamma \ll 1$, but having infinitely many of them contributing as $s \to s_0$. 

Concretely, we assume the partial wave to assume the form \eqref{eq:breitwigner} up to some spin $L$, above which we assume no further interaction, so that $S_{\ell > L} = 1$. The cross-section close to $s \sim s_0$ is then given by the finite sum
\be
\sigma_\text{tot}(s \sim s_0) \propto \sum_{\ell = 0}^L n_\ell^{(d)} {2 \Gamma^2 \over (s - s_0)^2 + \Gamma^2} 
\ee
where $n_\ell^{(d)}$ is given below \eqref{eq:T_partial_waves}. Since at large $\ell$ we have $n_\ell^{(d)} \sim \ell^{d-3}$ we find for sufficiently large $L$
\be
\sigma_\text{tot}(s \sim s_0) \propto {  L ^{d-2} \, \Gamma^2 \over (s - s_0)^2 + \Gamma^2} 
\ee
Taking $\Gamma \sim \epsilon$ and $L^{d-2} \sim 1/\epsilon$, in the limit $\epsilon \to 0$ we find a point-wise divergent cross-section
\be
\sigma_\text{tot}(s\sim s_0) \propto \delta(s - s_0)
\ee
Since the delta-function is integrable, the upper bound on the integrated cross-section $\bar{\sigma}_\text{tot}(s > s_0)$ can still be respected.

We can indeed verify this numerically by directly maximizing the cross-section point-wise with the S-matrix Bootstrap. In Figure \ref{fig:max_pointwise}, we show the maximal value of $\sigma_\text{tot}(60)$ as a function of $N$. We find that the objective grows almost linearly in $N$ not showing any sign of convergence. In the same Figure, we observe a surprising universal feature: the profile of the cross-section saturating the bound on $\sigma_\text{tot}(60)$ for $N=20$ is almost identical to the profile of the Froissart Amplitude \ref{fig:Froissart_total_sigma} at energies $\sqrt{s}\ll \sqrt{60}$. This universality further supports the notion that maximizing the cross-section at a given energy requires the excitation of all spins at that scale: a goal achievable only via Reggeization of these resonances, which enhances the interaction at lower energies.

Note that it is also possible to have an accumulation of resonances in the physical region that makes the cross-section highly oscillatory (but still finite).  This phenomenon is known as \emph{Martin's pathology} \cite{Martin:933379} and, so far, has not been ruled out by basic principles.

 \section{Froissart bound in $d$-dimensions}
 \label{sec:Analytical_bound}
In this Appendix, we  discuss the asymptotic behaviour of the bound in \eqref{eq:sigmatotanalytic}.
We will interpret this limit as the $d$-dimensional version of the Froissart bound. Similar derivations can also be found in \cite{sasha, Haldar:2019prg}.

The partial waves in $d$ dimensions admit the integral representation
\be
\mathcal{P}_\ell^{(d)}(z){=}\frac{\Gamma(\tfrac{d}{2}-1)}{\sqrt{\pi}\Gamma(\tfrac{d-3}{2})}\int\limits_0^\pi (z+\sqrt{z^2{-}1}\cos\phi)^\ell \sin^{d{-}4}\phi d\phi.
\ee
From the definition, it immediately follows that \footnote{We use the fact that the integrand is positive, and that the $(z+\sqrt{z^2-1}\cos\phi)$ function is monotonically decreasing.} -- see also \cite{Haldar:2019prg}
\be
\mathcal{P}_\ell^{(d)}(z)\geq \frac{\Gamma(\tfrac{d}{2}-1)}{\sqrt{\pi}\Gamma(\tfrac{d-3}{2})} (z+\sqrt{z^2{-}1}\cos\phi_0)^\ell \mathbf{k}(\phi_0),
\ee
where $\mathbf{k}(\phi_0)=\int\limits_0^{\phi_0} \sin^{d-4}\phi d\phi$. 
Using the above inequality, it is easy to inspect the leading asymptotic behavior of the bound function $\Sigma_d$ given in \eqref{eq:sigmatotanalytic}, when we take both large $s$ and large $L$ 
\bea
\label{eq:asymptotic_to_minimize}
s^{\tfrac{d}{2}-2}\Sigma_d(s,t_0,k,L)\sim \mathbf{a}L^{d-2}+\mathbf{b} \mathbf{c}^{-L},
\eea
where
\bea
&&\mathbf{a}=\frac{2^{4-d}}{\Gamma(\frac{d}{2})\Gamma(\tfrac{d}{2}-1)},\quad \mathbf{c}=1 + 2 \sqrt{\frac{t_0}{s}} \cos\phi_0 \nn\\
&&\mathbf{b} = {2^k \pi^{3 \over 2} \Gamma({d - 3 \over 2})  \over  \Gamma({d \over 2} - 1) \mathbf{k}(\phi_0)} \,c_{2k}(t_0) \,s^{2k + {d - 4 \over 2}}\nn
\eea
The equation \eqref{eq:asymptotic_to_minimize} has a minimum at 
\be
L^*\sim (2k-1)\sqrt{\frac{s}{t_0}}\frac{\log(s)}{2\cos\phi_0}.
\ee
Evaluating the bound for $L=L^*$, taking the leading large $s$ term and setting $\phi_0=0$ yields the asymptotic behaviour of the bound function
\be
\label{eq:asymptotic_bound}
\Sigma_d(s,t_0,k,L^*)\simeq \frac{(2k-1)^{d-2}}{4^{d-3 }\Gamma(\tfrac{d}{2})\Gamma(\tfrac{d}{2}-1)} \, {s \log^{d-2} s \over t_0^{{d \over 2}-1}}
\ee
If we wish to further minimize the above expression we can take $k=1$, and $t_0=4$. With this choice, we recover the Froissart bound \eqref{eq:froissart_in_d_dims}.
The bound in \eqref{eq:asymptotic_bound} implies the Regge bound on the amplitude $T(s,t_0)< C s^2$ for $s$ real, and 
by the \emph{Phragmen-Lindelof theorem} we can 
extend this condition on the arc at infinity. 
The Regge bound permits writing double-subtracted dispersion relations for the amplitude, which is consistent with the $c_2(t_0)$ coefficient being dispersive.

\vspace{3mm}

\noindent \emph{Massless particles.} Discussing the Froissart bound in the context of massless particles requires extra care. The naive claim that long-range interactions invalidate the Froissart bound should be refrased: the bound cannot be derived when there is no consistent non-perturbative definition of the scattering amplitude.

Here we will briefly discuss an example in which the amplitude of massless particles makes sense non-perturbatively: the case of pions or Goldston amplitudes. In this case, we imagine an amplitude that at low energy behaves like $T(s,t)=\tfrac{g_2}{f^4} (s^2+t^2+u^2)+\dots$, and that satisfies the Regge bound $\lim_{s\to\infty}T(s,0)s^{-2}=0$. In this case, taking the limit $m\to 0$ in equation \eqref{eq:sigmatotanalytic} at fixed decay constant $f$ yields the rigorous bound
\be
s^{\tfrac{d}{2}-2}\bar\sigma_\text{tot}(s)\leq \frac{\mathcal{N}_d 2^d \pi^{\tfrac{d}{2}-1}}{\Gamma(\tfrac{d}{2})}(L-1)_{d-2}+\frac{2\pi g_2}{f^4} s^{\tfrac{d}{2}}.
\ee
Since the right hand side does not depend on $L$ anymore -- when particles are massless $t_0=0$ -- minimization in $L$ is trivial and sets $L=0$ giving
\be
\label{eq:a_anomaly}
\bar\sigma_\text{tot}(s)\leq \frac{2\pi g_2}{f^4}s^2.
\ee

If we consider the scattering of dilatons in four dimensions, then $2\pi g_2=\Delta a$. This simple observation relates the $a$-anomaly coefficient to the universal bound on the cross-section. The bound \eqref{eq:a_anomaly} only holds if we know the coefficient $g_2$. Unlike the gapped case, here $\max g_2=\infty$, and we cannot derive a universal bound for massless particles. 

\section{Regge Theory and Froissart growth}
\label{sec:regge}

\noindent In this appendix we comment on the implications of Froissart growth in the context of Regge theory, with particular focus on the Pomeron trajectory. We start by introducing the Sommerfeld-Watson representation and its use in Regge theory. Please refer to \cite{Gribov:2003nw, Collins:1977jy} for a more comprehensive introduction (or Appendix A in \cite{Albert:2023jtd} for a modern text). 

Regge theory is based on analyticity in the complex angular momentum $\ell$ plane.\footnote{This assumption is referred to as \emph{maximal analyticity of the second kind} in the old literature \cite{Collins:1977jy}. It has not been proven from axiomatic QFT.} Under this assumption we can re-write the partial wave decomposition \eqref{eq:T_partial_waves} as a contour integral in $\ell$. In particular, in the crossed-channel ($s \leftrightarrow t$) we have 
\be
\label{eq:SW}
T(s,t) = {1 \over 2 i} \oint_C n_\ell^{(d)} {\mathcal{P}^{(d)}_\ell(-\bar{z}) \over \sin \pi\ell} f_\ell(t) \, d\ell
\ee
where the contour $C$ encircles every integer anti-clockwise, with $\bar{z} = z|_{s \leftrightarrow t} = 1 + {2 s \over t - 4}$ and $f_\ell(t) \equiv {\sqrt{t} \over  (t - 4)^{d -3 \over 2} } \big( 1 - S_\ell(t) \big)$ is the \emph{connected} partial wave amplitude in the $t$-channel. 
Note that $n_\ell^{(d)}$ given below \eqref{eq:T_partial_waves} admits a good analytic continuation in $\ell$. 

\begin{figure}[h!]
    \centering
    \includegraphics[width=\linewidth]{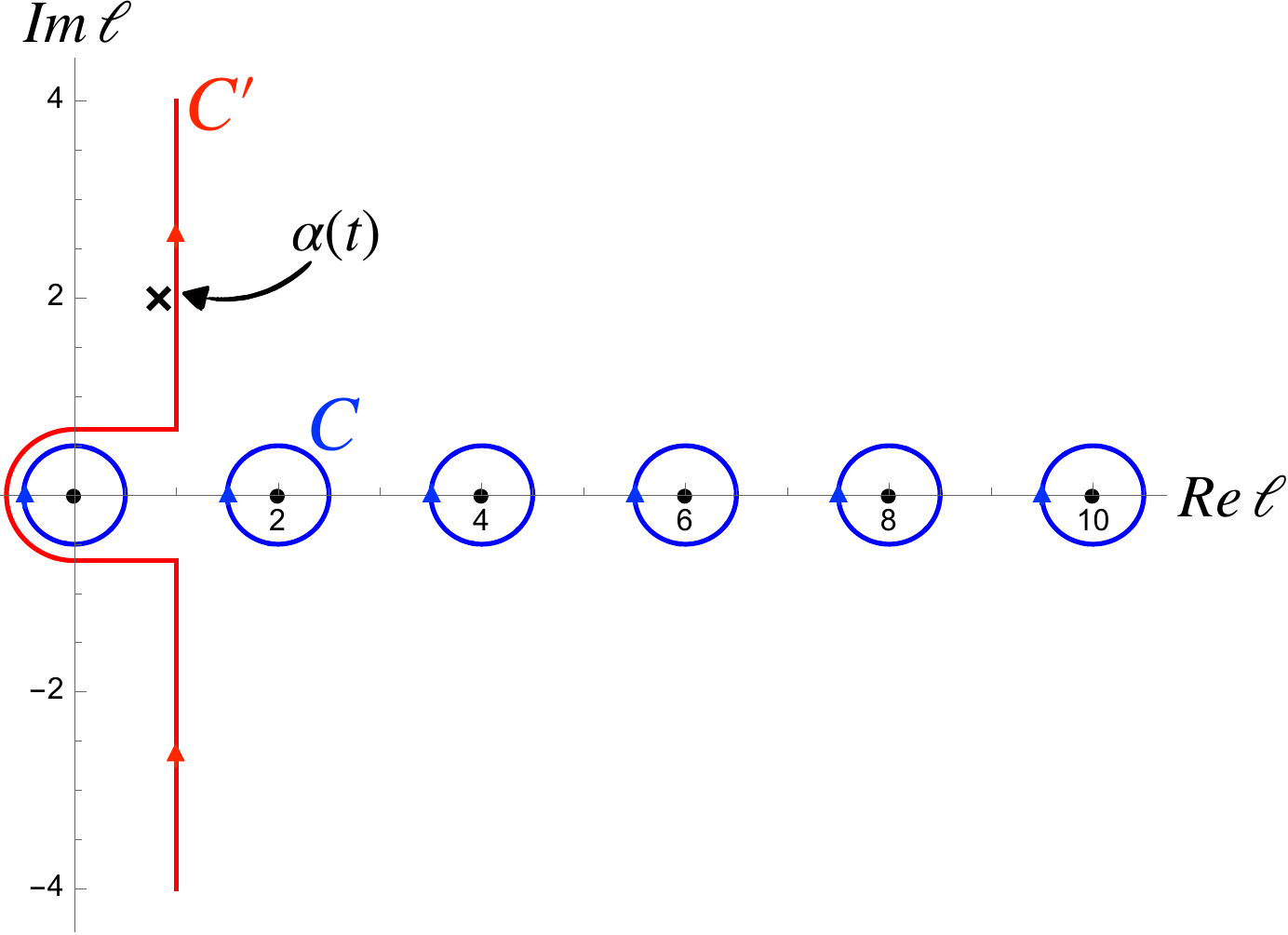}
    \caption{Complex angular momentum plane. In blue, the initial contour over even integer spins corresponding to the partial wave decomposition. In red, the deformed contour in the Sommerfeld-Watson representation.}
    \label{fig:complex_ell_plane}
\end{figure}

Now, we use the exponential decay of $f_\ell(t)$ in $\ell$ to open the contour $C$ from the right and align it on the imaginary axis for some $\Re \ell = c > \Re{\alpha(t)}$ where $\ell =\alpha(t)$ is the right-most singularity of $f_\ell(t)$. Dropping the arcs we end up with the contour $C'$ in Figure \ref{fig:complex_ell_plane}.\footnote{We have glossed over several important details regarding this analytic continuation (see \cite{Collins:1977jy,Gribov:2003nw} for proper detail).} 

Along contour $C'$, expression \eqref{eq:SW} is known as Sommerfeld-Watson representation and, contrarily to the partial wave decomposition \eqref{eq:T_partial_waves}, it can be used in both $s$- and $t$-channels.  In particular, we are interested in the $s$-channel high-energy regime  $s \to \infty$ where $\mathcal{P}_\ell^{(d)}(-\bar{z}) \sim (- \bar{z})^\ell \sim s^\ell$. In this limit the Sommerfeld-Watson representation reduces to an inverse Mellin transform:\footnote{Conversely, the tail of the Froissart-Gribov integral \eqref{eq:FG}, which determines the singular behavior in $\ell$, takes the form of a Mellin transform owing to $Q_\ell(\bar{z}) \sim s^{-\ell-1}$ as $s \to \infty$ (where we crossed $s \leftrightarrow t$).}
\be
\label{eq:invmellin}
T(s \to \infty,t) \simeq {1 \over 2 \pi i} \int\limits_{c - i \infty}^{c + i \infty} s^\ell \, \tilde{f}_\ell(t) \, d \ell
\ee
where we absorbed all $s$-independent factors into $\tilde{f}_\ell(t) \propto f_\ell(t)$. Their precise form will not concern us here.
\par
Suppose that the rightmost singularity is a Regge pole at $\ell \to \alpha(t)$ of order $n$,
\be
\tilde{f}_\ell(t) \sim {1 \over \big( \ell - \alpha(t) \big)^n}.
\ee
Sliding the contour to the left and picking up the residue across this pole lets us determine the leading behavior at large $s$,
\be
\label{eq:Froipom}
T(s \to \infty, t) \sim s^{\alpha(t)} \log^{n-1} \! s.
\ee
We thus conclude that in order to have Froissart bound saturation $T(s \to \infty,0) \sim s \log^{d -2} s$ we must have a Regge pole of order $n = d -1$ with intercept $\alpha(0) = 1$. In $d = 4$ this corresponds to a third-order Regge pole as mentioned in the text.

The rightmost Regge trajectory that is responsible for cross-section growth is typically referred to as the Pomeron. Let us now show that Froissart growth \emph{cannot} be generated by a linear Pomeron trajectory $\alpha(t) = 1 + \alpha' t$. The original argument is due to Oehme \cite{Oehme:1971omw,Oehme1972}, which we summarize here in general $d$-spacetime dimensions.

The basic observation is that if the total cross-section $\sigma_{tot}(s) \sim \Im T(s,0)/s $ grows, then the elastic differential cross-section in the forward limit will grow even faster because it involves the square of the amplitude,
\be
{d \sigma_{el} \over dt}\Big|_{t \to 0} \sim  {|T(s,0)|^2 \over s^2} \gg \sigma_{tot}(s).
\ee
On the other hand, unitarity implies
\be
\label{eq:unieltot}
\sigma_{el}(s) \equiv \int_{-s}^0 {d\sigma_{el} \over dt} dt \leq \sigma_{tot}(s).
\ee
The above two conditions can only be satisfied if ${d \sigma_{el} \over dt}$ sharply decays away from the forward limit $t < 0$. Moreover, the rate of decay of ${d \sigma_{el} \over dt}$ \emph{must} increase with energy. These observations qualitatively explain why the elastic differential cross-section should have a ``shrinking cone" as seen in Figure \ref{fig:diffractive_cone} for small $|t| \lesssim 0.5$.

The differential cross-section for the linear Pomeron model \eqref{eq:Froipom} reads 
\be
\label{eq:Froipom2}
{d \sigma_{el} \over dt} \sim s^{2 \alpha' t} \log^{2d-4} s
\ee
and is expected to hold in a neighborhood of $t \sim 0$. From \eqref{eq:unieltot} and since ${d \sigma_{el} \over dt} > 0$ the weaker bound follows
\be
\label{eq:difconebound}
\int^0_{-t_0} {d \sigma_{el} \over dt} dt \leq \sigma_{tot}(s),
\ee
for sufficiently small but finite $t_0$, that is $-s\ll t_0 < 0$. Plugging \eqref{eq:Froipom2} into the above, we see that $\int_{-s_0}^0 s^{2 \alpha' t} dt \sim 1/\log s$ in the limit $s \to \infty$ which indeed provides some shrinking and makes the LHS of \eqref{eq:difconebound} grow as $\log^{2 d - 5} \! s$. However the RHS goes as $\log^{d-2} \!s$ and the bound is violated for $d \geq 4$.

This forces one to move away from the simple linear Pomeron trajectory picture. As Oehme suggests \cite{Oehme:1971omw}, one way out is to have a Pomeron trajectory approaching the intercept $t \to 0$ with \emph{infinite} slope, such as $\alpha(t) = 1 + r_0 \sqrt{t}$. This is precisely how the white ring diffraction model is able to achieve Froissart growth in a way compatible with unitarity away from the forward limit (see Appendix \ref{sec:whitering}).

The numerical S-matrix Bootstrap, which crucially also accounts for full non-perturbative unitarity in the $t$-channel, appears to present an even more complicated mechanism, as seen from the Chew-Frautschi plot in Figure \ref{fig:chew_frauschi}, where the Regge trajectories  $A'$ and $Z_i$ have similarly steep slopes near $t \sim 0$.

\section{White Ring diffraction model}
\label{sec:whitering}

\noindent Here we derive a ring diffraction model for the scattering amplitude. The starting point is the eikonal representation for the amplitude (see e.g. \cite{Bellazzini:2022wzv,Haring:2024wyz}). At large enough $s$ we have
\be
\label{eq:eikonal}
T(s,t) \simeq 2 i s \int d^{d-2} \mathbf{b} \, e^{- i \mathbf{q} \cdot \mathbf{b}} \big(1 - S(s,b) \big)
\ee
with $t = - |\mathbf{q}|^2$ and $b = |\mathbf{b}|$ is the impact parameter,
where $S(s,b)$ relates to the partial wave amplitude $S_\ell(s)$ via $\ell \simeq  b \, \sqrt{s}/2$ at large $s$. In particular, $S(s,b)$ satisfies unitarity $|S(s,b)|^2 \leq 1$.

Switching to spherical coordinates in \eqref{eq:eikonal} and integrating over the angles leads to\footnote{This formula can be derived directly from the partial wave decomposition \eqref{eq:T_partial_waves} by letting $\ell = b \sqrt{s}/2 \to \infty$ and keeping $b$ finite. In this limit the sum over even $\ell$ becomes $\sum_\ell \to {\sqrt{s} \over 4} \int db $, and also $n_\ell^{(d)} \to \ell^{d-3}\,  2^{d+1} \pi^{d -2\over 2} / \Gamma({d-2 \over 2})$. Finally, using the identity $ \mathcal{P}_\ell^{(d)}(1 -{x \over 2 \ell^2}) \simeq \Gamma({d - 2\over 2})  2^{d-4 \over 2} x^{4 - d \over 4} J_{d -4 \over 2}(\sqrt{x})$ at large $\ell$ (from eqs. (8.1.2) and (9.1.71) of \cite{abramowitz1965handbook} and definition of $\mathcal{P}_\ell^{(d)}$ in terms of $_2 F_1$ from eq. (2.31) in \cite{Correia:2020xtr}) we arrive at \eqref{eq:eikonal2}.  
}
\be
\label{eq:eikonal2}
\!T(s,t) \simeq 2 i s (2 \pi)^{d -2  \over 2} \!\!\int\limits_0^\infty \!db \,{b^{d-2 \over 2} \over |\mathbf{q}|^{d -4 \over 2}} J_{d-4 \over 2}(b |\mathbf{q}|)  \big(1 - S(s,b) \big),
\ee
where $J_{d - 4 \over 2}$ is the Bessel function of the first kind.

We consider a ``white ring" model which assumes complete reflection for $R_1 \leq b \leq R_2$, and no interaction otherwise, where $R_1$ and $R_2$ are respectively the inner and outer radii of the ring. Concretely, this amounts to the eikonal profile
\be
\label{eq:ringprofile}
\!\!\text{White Ring: } \; S(s,b) = 
\begin{cases}
+1   & \text{if }b <R_1 \\
-1 & \text{if } R_1\leq b \leq R_2 \\
+1   & \text{if } b >R_2 
\end{cases}\!\!
\ee
Performing the integral \eqref{eq:eikonal2} 
 with this profile leads to the amplitude for white ring diffraction:
\begin{align}
\label{eq:whitering}
\!T(s,t) = {4 i s (2 \pi)^\gamma \over(-t)^{\gamma \over 2}} &\Big[ (R_2)^\gamma \,J_\gamma(R_2 \sqrt{-t}) \nn
\\ &\;\;\;\;-  (R_1)^\gamma\, J_\gamma(R_1 \sqrt{-t})\Big] 
\end{align}
with $\gamma \equiv d/2 -1$.

The total cross-section is obtained from $\sigma_{tot}(s) \simeq \Im T(s,0)/s$, giving
\be
\label{eq:whiteringcs}
\sigma_{tot}(s) = {4 \pi^{{d \over 2} - 1} \over \Gamma({d \over 2})} \Big[(R_2)^{d -2 } - (R_1)^{d -2 } \Big]
\ee
which is four times the volume of a ring in $d-2$ spatial dimensions. The elastic differential cross-section of the white ring is obtained by inserting \eqref{eq:whitering} into \eqref{eq:elasticsigma}. 

We observe that Froissart growth $\sigma_{tot}(s) \sim \log^{d-2}s$ is achieved if the ring grows logarithmically with energy, $R_2 \sim \log s$.\footnote{The inner radius $R_1$ may also evolve as long as it does not grow faster than $R_2$.} In Appendix \ref{sec:regge}, we concluded that unitarity rules out the possibility of Froissart growth coming from a linear Regge trajectory. However, the white ring model is built to satisfy unitarity by design, as shown in Eq.~\eqref{eq:ringprofile}. So, the question now is: what kind of Regge trajectory could lead to Froissart growth in the white ring model?

It suffices to consider the outer radius $R_2$ contribution in \eqref{eq:whitering} (the $R_1$ term will either contribute in a similar way or drop out in the limit $s \to \infty$). Let $R_2(s) = r_0 \log s$ where $r_0$ is a dimensionful scale satisfying $0<r_0 \leq 1/  \sqrt{t_0}$, according to the bound \eqref{eq:froissart_in_d_dims} and \eqref{eq:whiteringcs}, where $t_0$ is the nearest singularity in the $t$-channel. 

We make use of the following integral representation of the Bessel function \cite{Arfken:379118}\footnote{This representation is valid for $\gamma$ integer. Since $\gamma = {d \over 2} - 1$ our argument will only apply in even dimensions. The odd dimensional case should be simpler since the Bessel function is given in terms of elementary functions.}
\be
J_\gamma(x) = {1 \over 2\pi i} \oint_{j=0} e^{x(j - j^{-1})/2} \, j^{-\gamma - 1} \, dj
\ee
where the contour encircles $j = 0$ counter-clockwise. Under the change of variable $\tilde{\ell} = (j - j^{-1})/ 2 i$ we get
\be
J_\gamma(x) = {1 \over 2\pi i} \! \oint\limits_{(-1,1)} \!\! e^{i x \tilde{\ell}} \; {\big(i \tilde{\ell} + i \sqrt{\tilde{\ell}^2 - 1}\big)^{- \gamma} \over \sqrt{\tilde{\ell}^2 - 1}} \, d\tilde{\ell}
\ee
where the contour encircles the branch-cut $\tilde{\ell} \in (-1,1)$ counter-clockwise. Given $x = R_2 \sqrt{-t} = r_0 \log s\, \sqrt{-t}$ we have for the top line of \eqref{eq:whitering}
\begin{align}
\label{eq:Reggeint}
T(s&,t) \supset {4 i (2 \pi r_0)^\gamma \log^\gamma s \over 2\pi i\,  (-t)^{\gamma \over 2}} \times \nn \\
& \oint\limits_{(-1,1)} \!s^{1+ i r_0 \sqrt{-t} \, \tilde{\ell}} \;{\big(i \tilde{\ell} + i \sqrt{\tilde{\ell}^2 - 1}\big)^{- \gamma} \over \sqrt{\tilde{\ell}^2 - 1}} \, d\tilde{\ell}
\end{align}
Finally, we change variable to $\ell = 1 + i r_0 \sqrt{-t} \, \tilde{\ell}$ and use the relation ${d^\gamma \over d\ell^\gamma}  s^\ell= (\log^\gamma \!s)\,   s^\ell  $ to find 
\begin{align}
&T(s,t) \supset {4 i (- 2 \pi r_0^2)^\gamma \over 2 \pi i} \times \nn \\
& \;\oint_F d\ell\, s^\ell \,{d^\gamma \over d\ell^\gamma} \bigg[{\big(\ell-1 + \sqrt{(\ell-1)^2 - r_0^2 t} \big)^{- \gamma} \over \sqrt{(\ell-1)^2 - r_0^2 t}} \bigg]
\end{align}
where the $F$ contour is depicted below
\begin{figure}[H]
    \centering
    \includegraphics[width=0.45\linewidth]{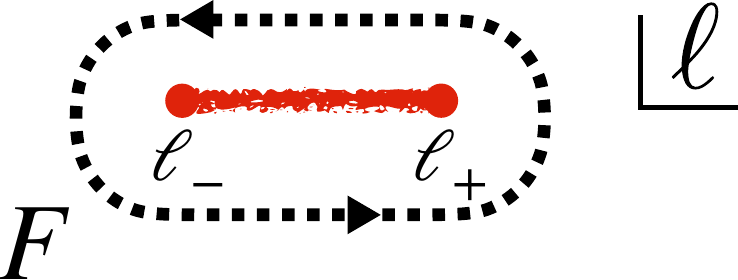}
    \label{fig:enter-label}
\end{figure}
\noindent encircling a branch-cut with end-points
\be
\label{eq:lpm}
\ell_{\pm} = 1 \pm r_0 \sqrt{t}.
\ee
Note that equation \eqref{eq:Reggeint} is precisely the inverse Mellin transform in eq. \eqref{eq:invmellin} allowing us to read off the partial wave $\tilde{f}_\ell(t)$ in the crossed-channel. 

In particular, we see that our model achieves Froissart growth via a Regge \emph{cut}: The $t$-channel partial wave $f_\ell(t)$ has a branch cut in $\ell$ with branch-points given in \eqref{eq:lpm}. In the forward limit $t \to 0$ we find that the branch-points \eqref{eq:lpm} coalesce, and the Regge cut degenerates into a Regge pole at $\ell = 1$:
\begin{align}
f_\ell(t) \sim {d^\gamma \over d\ell^\gamma} \bigg[{1 \over (\ell -1)^{1 + \gamma}} \bigg] \sim {1 \over (\ell- 1)^{1+2 \gamma}}.
\end{align}
From the relation $\gamma = {d \over 2} - 1$ we see that the Regge pole is of order $d - 1$, which is necessary for Froissart growth of the cross-section, as discussed below \eqref{eq:Froipom}.

The full expression for $f_\ell(t)$ in $d = 4$, or $\gamma = 1$, is given in \eqref{eq:pwt}.

\section{Summary of Numerical Parameters}
\label{sec:appendix_numerics_details}

In Section~\ref{sec:numerical_s_matrix_bootstrap}, we describe the overall numerical setup. In this Appendix, we specify the precise numerical parameters used to obtain the results presented in the paper.

We use eight different foliations:
\be
P = \left\{\tfrac{20}{3}\right\} \cup P' = \left\{ \tfrac{20}{3}, 10, 20, 30, 40, 50, 60, 86 \right\}.
\ee
The foliation with \( P = 20/3 \) corresponds to the original choice used in the first S-matrix bootstrap paper~\cite{Paper3}. The explicit form of the ansatz is given by
\be
T(s,t) = \mathcal{F}_{\tfrac{20}{3}}^N(s,t) + \sum_{p \in P'} \mathcal{F}_p^{N-2}(s,t).
\ee
For the largest value used, \( N = 20 \), the total number of ansatz parameters is \( |\alpha_{a,b}| = 814 \).

Unitarity is imposed on an adaptive grid of points that depends on the specific foliation choice. Defining
\be
\mathfrak{s}_p(\rho) = \frac{8(1 + \rho^2) - p(\rho - 1)^2}{(\rho + 1)^2},
\ee
the full unitarity grid is given by
\be
\mathcal{U}_\text{grid} =
\left( \bigcup_{p \in P'} \left\{ \mathfrak{s}_p(\rho) \,\middle|\, \rho \in \rho_\text{grid}^2 \right\} \right)
\cup
\left\{ \mathfrak{s}_{\tfrac{20}{3}}(\rho) \,\middle|\, \rho \in \rho_\text{grid}^1 \right\},
\ee
where the sets \( \rho_\text{grid}^1 \) and \( \rho_\text{grid}^2 \) are defined by
\be
\rho_\text{grid}^1 = \left\{ e^{i\phi} \,\middle|\, \phi \in \mathcal{G}_{300} \right\}, \quad
\rho_\text{grid}^2 = \left\{ e^{i\phi} \,\middle|\, \phi \in \mathcal{G}_{150} \right\},
\ee
and
\be
\mathcal{G}_n = \frac{\pi}{2} \left(1 + \cos\left( \frac{\pi k}{n + 1} \right) \right), \quad k = 1, \dots, n.
\ee

Unitarity is imposed for all partial waves up to \( \ell = L \equiv 18 \), via the condition
\be
|S_\ell(s)|^2 \leq 1, \quad s \in \mathcal{U}_\text{grid}.
\ee
The total number of unitarity constraints is \( |\mathcal{U}_\text{grid}| = 750 \). We have verified that decimating the grid (e.g., by retaining only one out of every four points) does not significantly affect the numerical results, and can be useful for accelerating the solution of the bootstrap optimization problem.

Using a higher spin cutoff \( L > 18 \) does not modify the results obtained with the values of \( N \) considered in this work.\footnote{
We remind the reader that the optimal solution is obtained in the double-scaling limit: first sending \( L \to \infty \) at fixed \( N \), and then taking \( N \to \infty \). It is therefore important to choose a sufficiently large value of \( L \) for each fixed \( N \).
}

As explained in Section~\ref{sec:numerical_s_matrix_bootstrap}, the inclusion of improved positivity constraints is essential for both the convergence of the $S$-matrix bounds and the stability of the optimal solution. These constraints (see Eq.~\eqref{eq:improved_pos_constraints}) are evaluated for all \( s \in \mathcal{U}_\text{grid} \) and for a dense set of values of \( t \in [0, 4) \), chosen using a grid that accumulates near \( t = 4 \).

\section{Amplitudes from direct maximization of $\bar{\sigma}_{tot}(s)$}

\begin{figure}
    \centering
    \includegraphics[width=\linewidth]{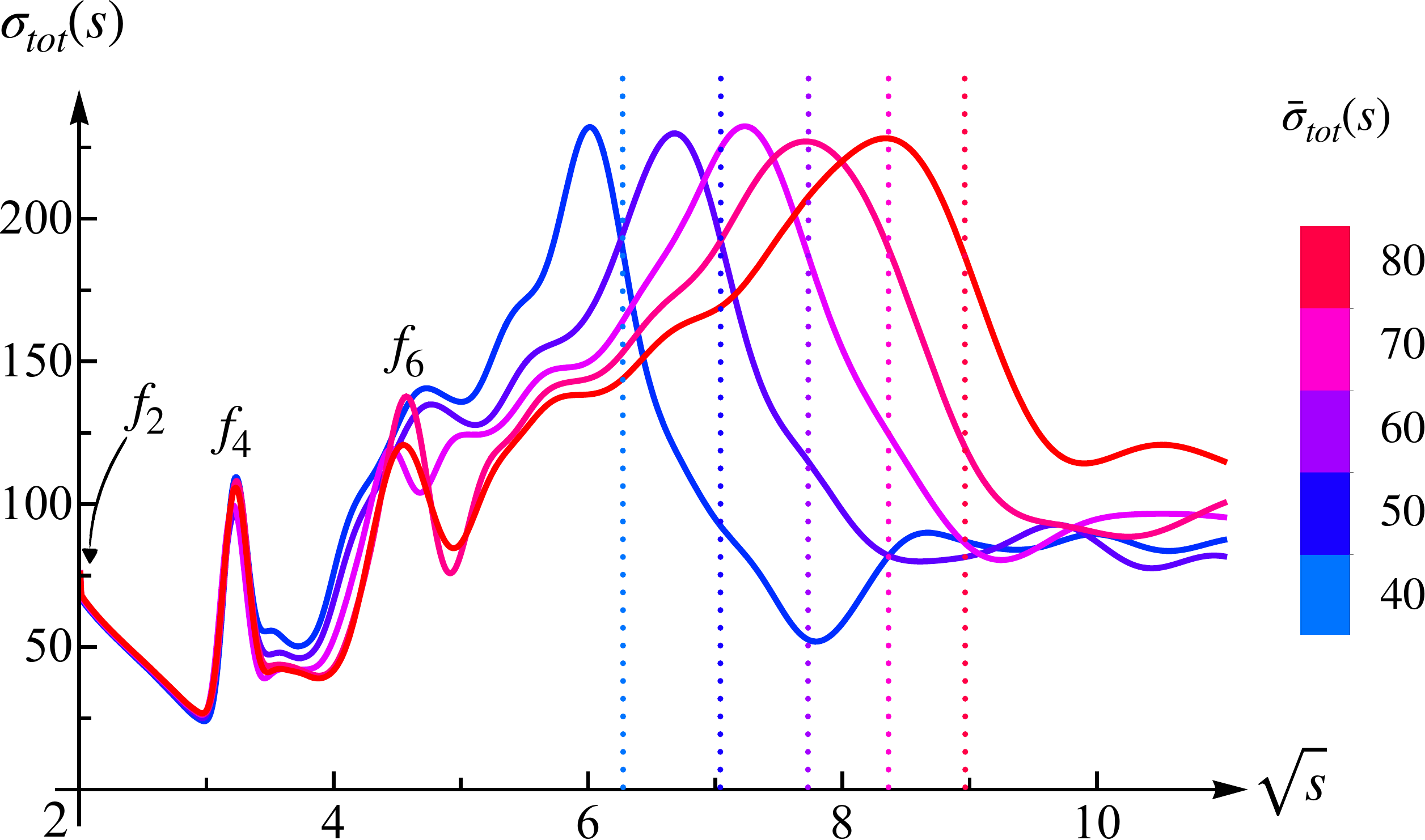}
    \caption{$\sigma_\text{tot}$ obtained by maximizing $\bar\sigma_\text{tot}(s)$ for different choices of $s$ (plotted with different colors). The window is also highlighted by the dotted vertical lines. At energy well below the window cutoff the profile is universal and it coincides with that of the Froissart amplitude.}
    \label{fig:yndurain_s}
\end{figure}

In this appendix we discuss in more detail the extremal amplitudes that maximize $\bar{\sigma}_{tot}(s)$. Unlike the Froissart amplitude, which is the universal solution to many different bootstrap problems. Here there is a different amplitude for each choice of $s$.

In figure \ref{fig:yndurain_s}, we plot the profile of $\sigma_\text{tot}(s)$ for the extremal solution of the optimization $\max \bar\sigma_\text{tot}(s)$.
This profile maximizes the integral of the cross-section in the window $[4,s]$. Though all profiles look different there is an interesting universality emerging. At low energy the profile is identical, then when the energy approaches $s$, we observe a pronounced peak. 

\begin{figure}
    \centering
    \includegraphics[width=\linewidth]{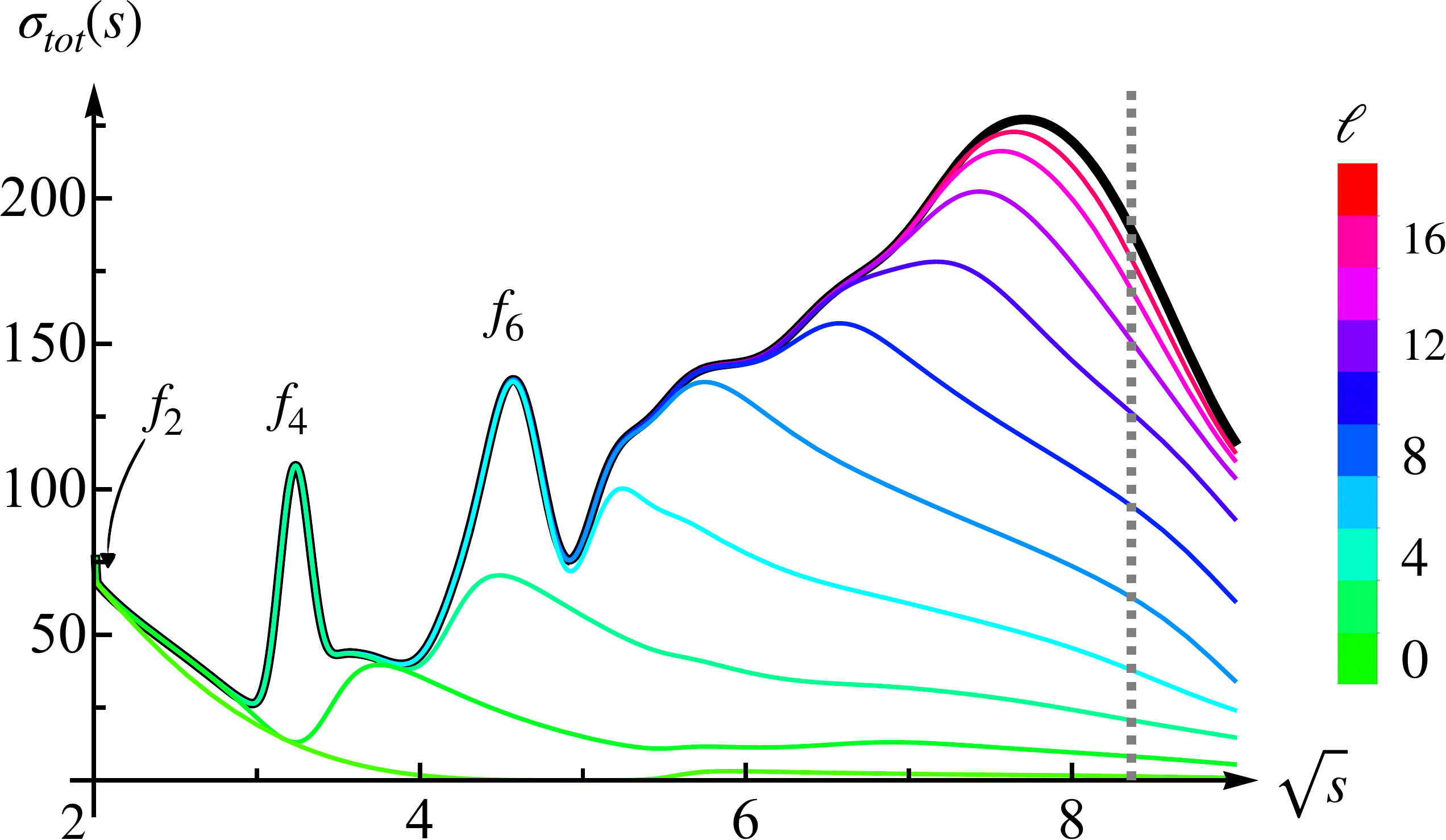}
    \caption{Profile of $\sigma_\text{tot}$ for the amplitude maximizing $\bar\sigma_\text{tot}(70)$ -- the integration window $[2,\sqrt{70}]$ is delimited by the dotted black line. We plot in different colours the sum of partial cross-sections up to spin $\ell$. We only show the results for $N=20$.}
    \label{fig:yndurain_at_70}
\end{figure}

Unlike the peaks in the low energy universal part which correspond to the $f_\ell$ resonances of the Froissart Amplitude, the peak close to $s$ is given by the sum all partial waves piling up to maximize the cross-section in the given window -- see Figure \ref{fig:yndurain_at_70}. Indeed, we can check that many resonances accumulate close to $s$, producing a trajectory reminiscent of the spectrum of a Coon amplitude \cite{Coon:1969yw}. This is the first time such an amplitude shows up in the non-perturbative $S$-matrix Bootstrap -- see \cite{Figueroa:2022onw} for a nice discussion in the context of meromorphic amplitudes.

The emergence of the Froissart Amplitude spectrum at low energy for the amplitudes maximizing the cross-section at finite energies as another indication of the conjecture presented in Section \ref{sec:max_cross_section}. We believe that if we were to send $s\to\infty$, the optimal amplitude would coincide with Froissart Amplitude at energies below $s$.\footnote{We also observe this universality for other choice of kernel we integrate over. As we push the integration domain high enough, we see similar peaks to the Froissart amplitude appearing at low energies.}

\begin{figure}
    \centering
    \includegraphics[width=\linewidth]{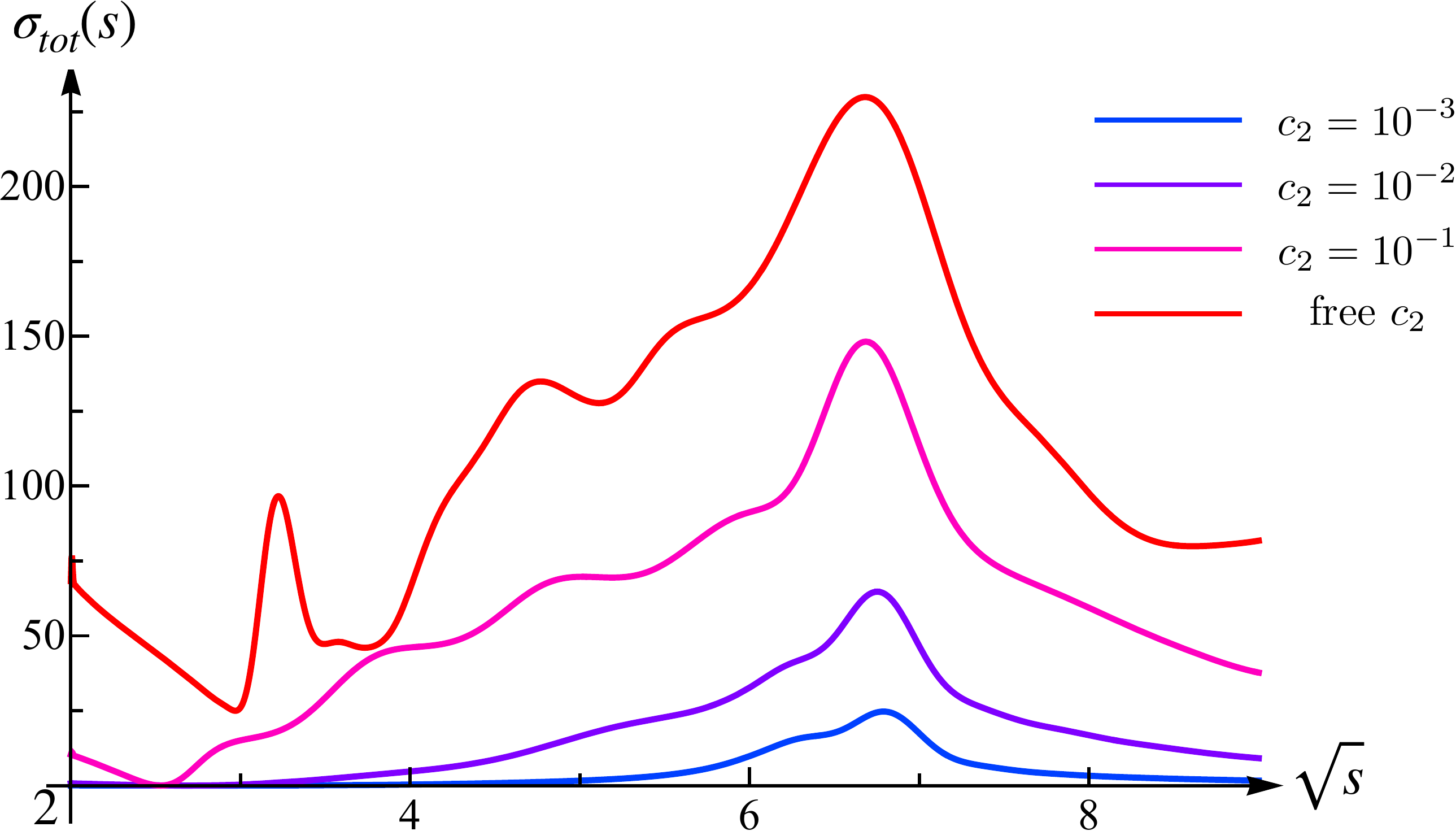}
    \caption{Profile of $\sigma_\text{tot}$ -- for the integration window $[2,\sqrt{70}]$ -- for different values of $c_2$. This plot showcase the role of $c_2$ as a "budget" for resonances, and the mechanism used by the S-Matrix bootstrap to maximize the cross-section by accumulating resonances close to the end of the integration window. We only show results for $N=20$.   }
    \label{fig:yndurain_fixed_c2}
\end{figure}

Lastly, we can explore how the optimal solution changes if we include an additional constraint on $c_2$. Fixing the value of $c_2$ should produce radically different amplitudes, since when $s\to\infty$ we could not approach the Froissart Amplitude anymore. In Figure \ref{fig:yndurain_fixed_c2}, we plot $\sigma_\text{tot}(s)$ from the amplitude maximizing $\bar\sigma_\text{tot}(70)$ with $c_2$ fixed at different values. We always observe a peak at the end-point of the integration window, but as expected, the low energy physics now it changes as we vary $c_2$. When $c_2$ becomes small, the amplitude profile resembles the profile of a weakly coupled amplitude below the cutoff $s$. As already discussed in \cite{EliasMiro:2022xaa, EliasMiro:2023fqi} $c_2$ can be interpreted as a \emph{budget} of resonances, that measures how strongly coupled the theory is at low-energy. This analysis further supports this idea, and it links again $c_2$ to the interaction strength and the Froissart bound.  

\section{Anatomy of the Froissart Amplitude}
\label{appendix:Froissart_amplitude}

\begin{figure}[h!]
    \centering
    \includegraphics[width=0.8\linewidth]{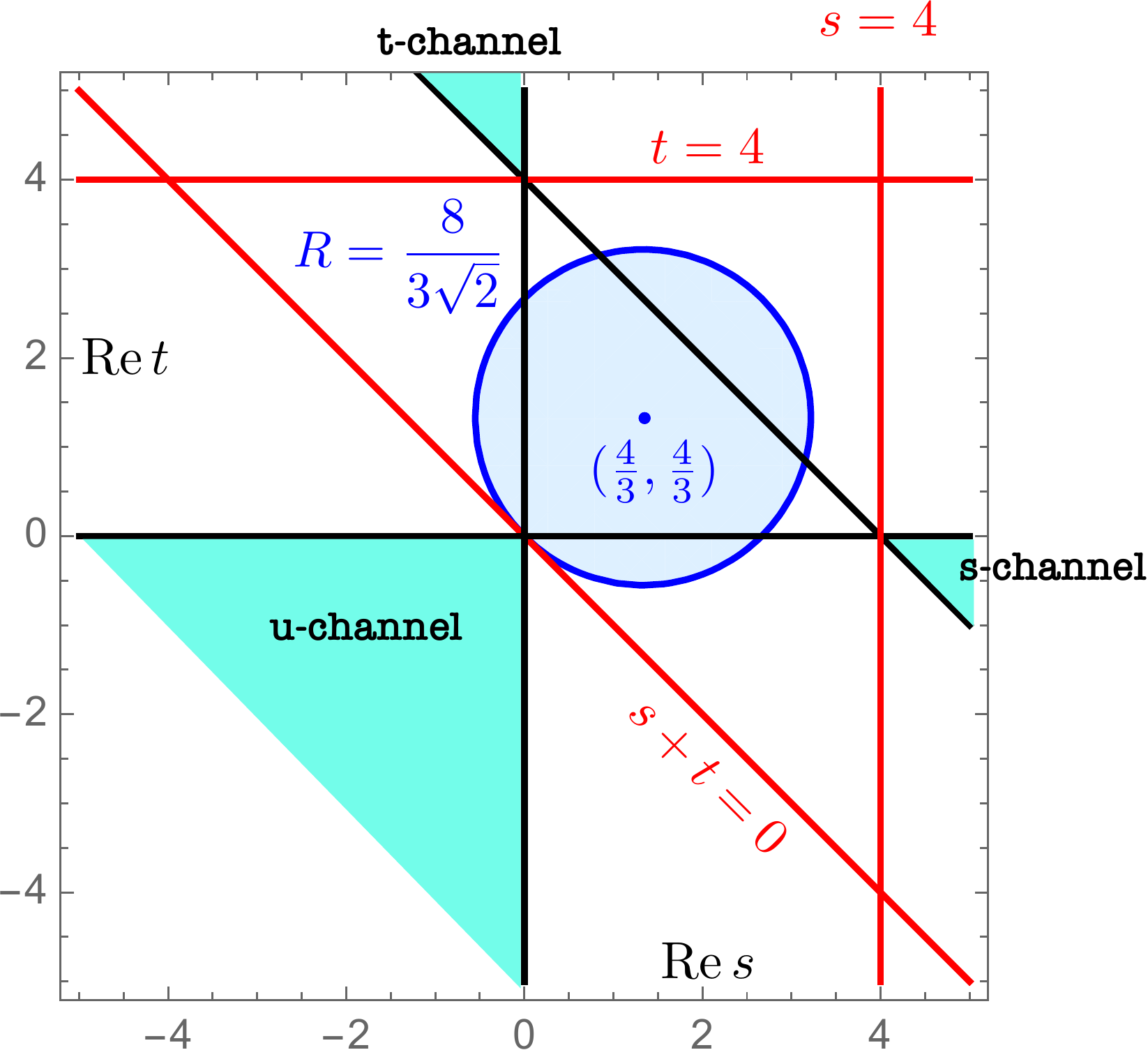}
    \caption{In blue, the convergence region of the Taylor expansion around the crossing-symmetric point. The black lines trace the physical regions associated with the \(s\)-, \(t\)-, and \(u\)-channels. In red, we show the projections of the singularity surfaces \(s = 4\), \(t = 4\), and \(s + t = 0\).}
    \label{fig:convergence_radius}
\end{figure}

A scattering amplitude can be characterized by its Taylor expansion coefficients around the crossing-symmetric point, denoted by \( \{c_{n,m}\} \), and commonly referred to as the \emph{S-matrix data}:
\be
T(s,t) = \sum_{n,m} c_{n,m} \, \sigma^n \tau^m,
\label{eq:S-matrix-data}
\ee
where \( \sigma = \bar{s}^2 + \bar{t}^2 + \bar{u}^2 \), \( \tau = \bar{s} \bar{t} \bar{u} \), and \( \bar{x} = x - \tfrac{4}{3} \) for \( x = s, t, u \). This expansion converges uniformly on any compact subset of the ball \( \| (s,t) - (\tfrac{4}{3}, \tfrac{4}{3}) \| < \tfrac{8}{3\sqrt{2}} \), with the radius of convergence set by the nearest singularity at \( s + t = 0 \). In Figure~\ref{fig:convergence_radius}, we plot the projection of the convergence domain of the expansion~\eqref{eq:S-matrix-data} onto the \( \{\Re s, \Re t\} \) plane.

Table~\ref{tab:s-matrix-data} reports the values of the S-matrix data extracted from the Froissart amplitude for \( N = 20 \), as well as their extrapolated values for \( N \to \infty \).

\begin{table}[h!]
    \centering
    \begin{tabular}{c|c|c}
        & \quad $N=20$ \quad & Extrapolation \\ 
    \hline
       $c_{0,0}$ & -7.05  & -7.3 \\
       $c_{1,0}$ & 0.929 & 0.97 \\
       $c_{0,1}$ & -1.81 & -1.9 \\
       $10^2\, c_{2,0}$ & 4.42 & 4.6 \\
       $10^2\, c_{2,1}$ & -4.81 & -4.9 \\
       $10^3\, c_{3,0}$ & 2.98 & 3.1 \\
       $10^2\, c_{0,2}$ & 3.71 & 3.7 \\
    \end{tabular}
    \caption{S-matrix data for the Froissart amplitude.}
    \label{tab:s-matrix-data}
\end{table}

\begin{figure}[t!]
    \centering
    \includegraphics[width=0.8\linewidth]{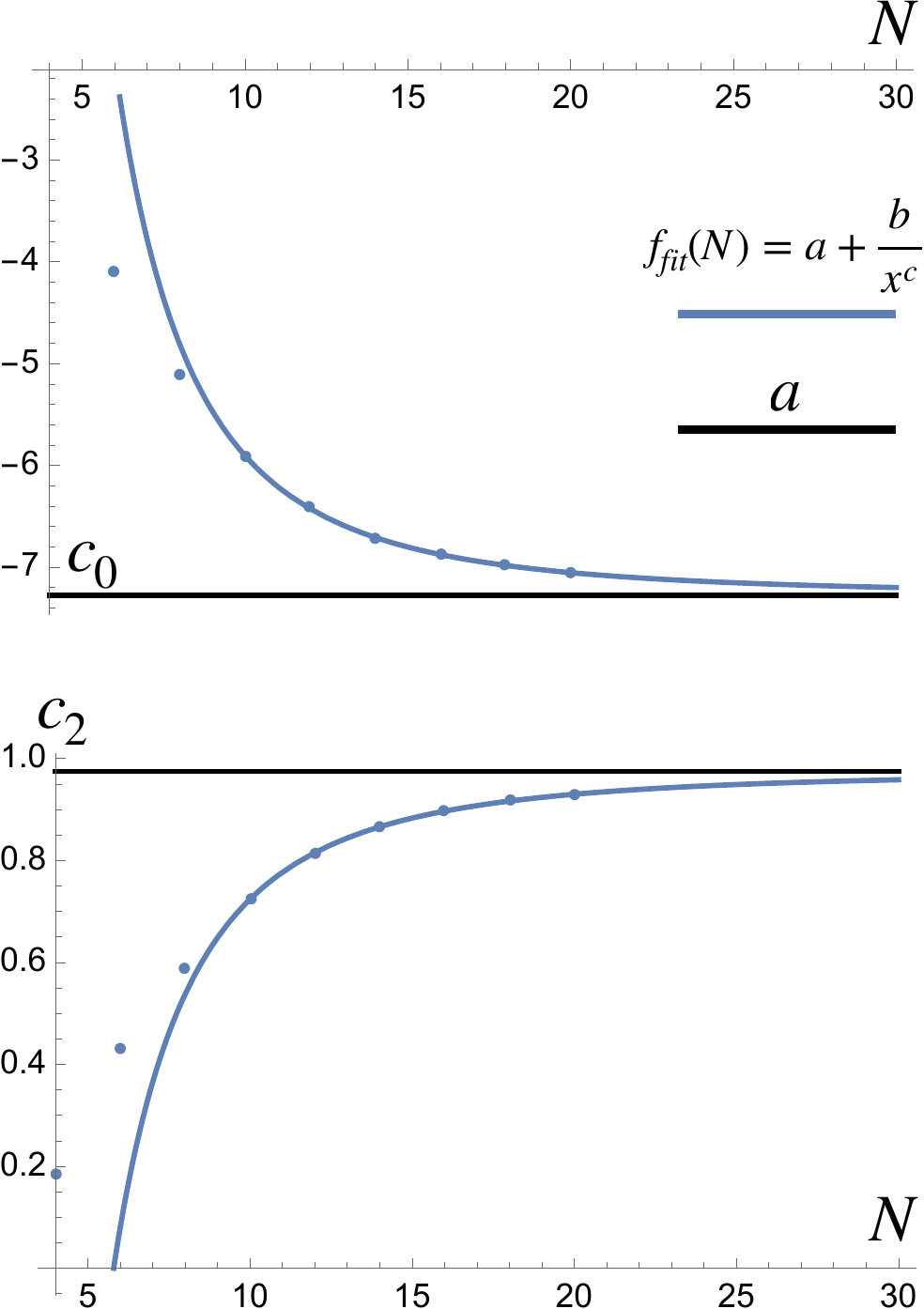}
    \caption{Behavior of \( c_0 \equiv c_{0,0} \) and \( c_2 \equiv c_{1,0} \) as functions of \( N \). Blue curves show best fits, while black markers denote extrapolated values. We use \( N \geq 10 \) for fitting, as lower \( N \) exhibits deviations from simple scaling behavior.}
    \label{fig:fits_wilson_coefficients}
\end{figure}

In Figure~\ref{fig:fits_wilson_coefficients}, we show how the leading S-matrix coefficients \( c_0 \equiv c_{0,0} \) and \( c_2 \equiv c_{1,0} \) approach their asymptotic bounds as \( N \to \infty \). Our results for \( c_0 \) are consistent with Figure 9 in~\cite{Paper3}, where the same quantity was computed using the original \( \rho \)-ansatz. Moreover, our extrapolations for \( c_0 \) and \( c_2 \) are compatible with the dual bounds obtained in~\cite{Guerrieri:2021tak, Gumus:2023xbs}.

Figure~\ref{fig:all_partial_waves} displays all partial waves up to \( \ell = 10 \) for the Froissart amplitude. For \( \ell \geq 4 \), the waves appear self-similar: they show a sharp resonance (defined by a phase shift passing through \( \pi/2 \)), followed by a plateau and a second, smaller jump. These features correspond to resonances on the Regge trajectories \( A \) and \( B \), associated with the \( f_\ell \) and \( g_\ell \) particles, respectively (see Figure~\ref{fig:chew_frauschi}).

The \( \ell = 0 \) and \( \ell = 2 \) partial waves behave differently. The \( \ell = 0 \) phase shift exhibits a maximally negative slope up to \( s \simeq 30 \), with a small jump thereafter. This behavior indicates a \emph{negative scattering length}. For \( N = 20 \), we obtain \( a_0 = -1.74 \), close to the rigorous bound \( a_0 \geq -1.75 \) from~\cite{Lopez:1975ca}.

\begin{figure}
    \centering
    \includegraphics[width=\linewidth]{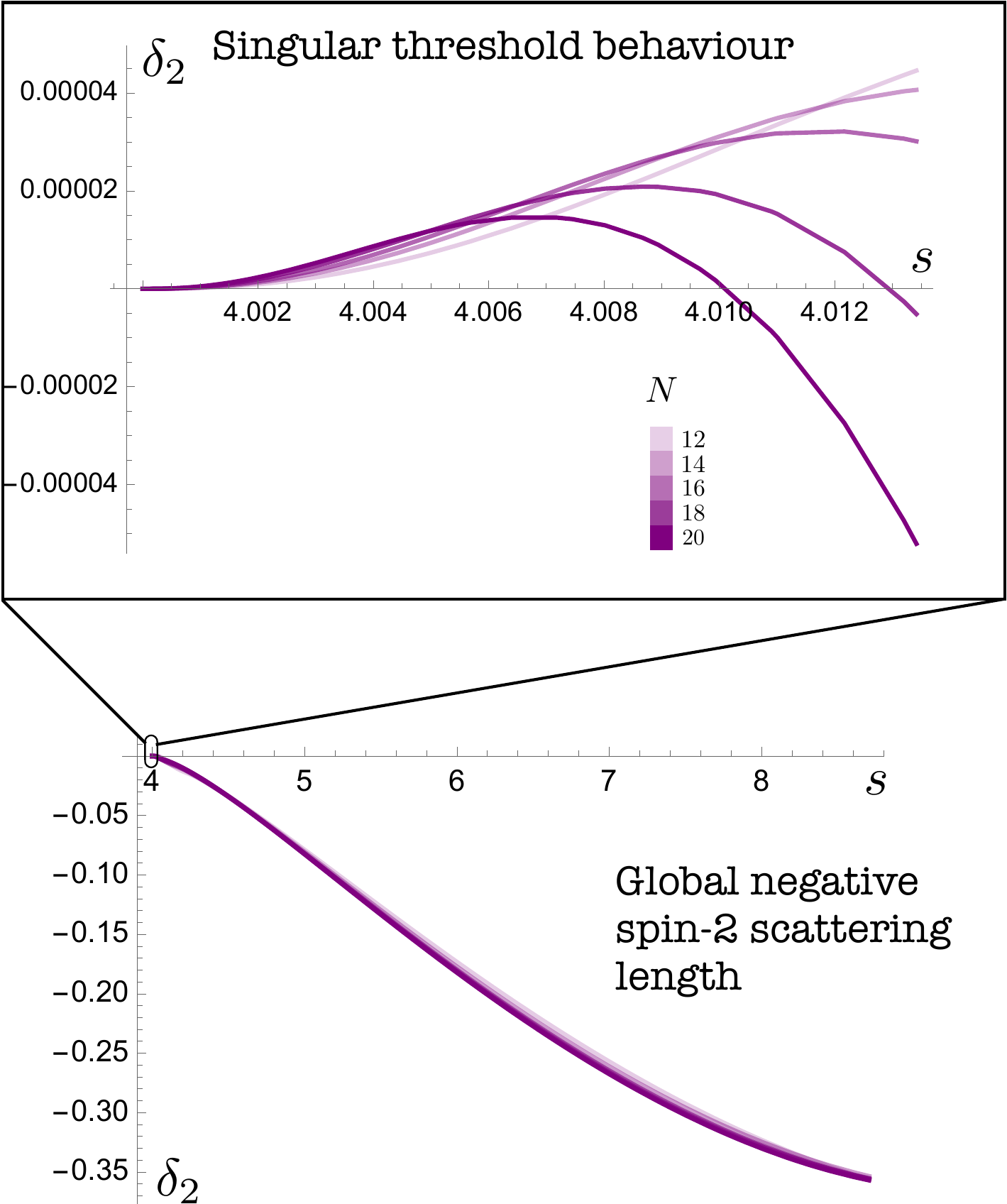}
    \caption{Phase shift for \( \ell = 2 \) as a function of \( s \). The top plot shows the threshold region, and the bottom plot displays a close-up around the mass scale. The increasingly singular behavior at small \( s \) leads to a divergent spin-2 scattering length as \( N \to \infty \).}
    \label{fig:singular_a2}
\end{figure}

The \( \ell = 2 \) partial wave also has a negative slope near threshold, followed by oscillations that converge slowly with \( N \). Although our ansatz has no singularity at \( s = 4 \), a zoomed-in view (Figure~\ref{fig:singular_a2}) reveals an increasingly sharp structure as \( N \to \infty \), with a steep rise in the slope followed by a rapid decrease. This suggests a \emph{growing spin-2 scattering length},
\be
a_2 = \frac{1}{32\pi^2} \int_4^\infty ds \, \frac{\Im T(s, 4)}{s^3} \geq 0.
\ee

To generate a negative spin-2 scattering length, the amplitude must admit a \emph{triple-subtracted dispersion relation at \( t = 4 \)}, requiring a spin-2 singularity at threshold. In~\cite{Guerrieri:2023qbg}, it was shown that the amplitude maximizing the residue of a spin-2 bound state with mass \( m_b^2 \) resembles the Froissart amplitude as \( m_b^2 \to 4^- \). Along the boundary of the allowed region (Figure~\ref{fig:c0c2}), instead, extremal amplitudes near the cusp exhibit spin-2 resonances with complex mass \( m_R^2 \to 4^+ \)~\cite{EliasMiro:2022xaa}.
Building on this evidence, we can confidently claim that the Froissart amplitude features a spin two threshold state.
Constructing a primal ansatz with such a singularity remains an open problem, as it would presumably be an amplitude presenting Froissart growth.

\begin{figure*}
    \includegraphics[width=\textwidth]{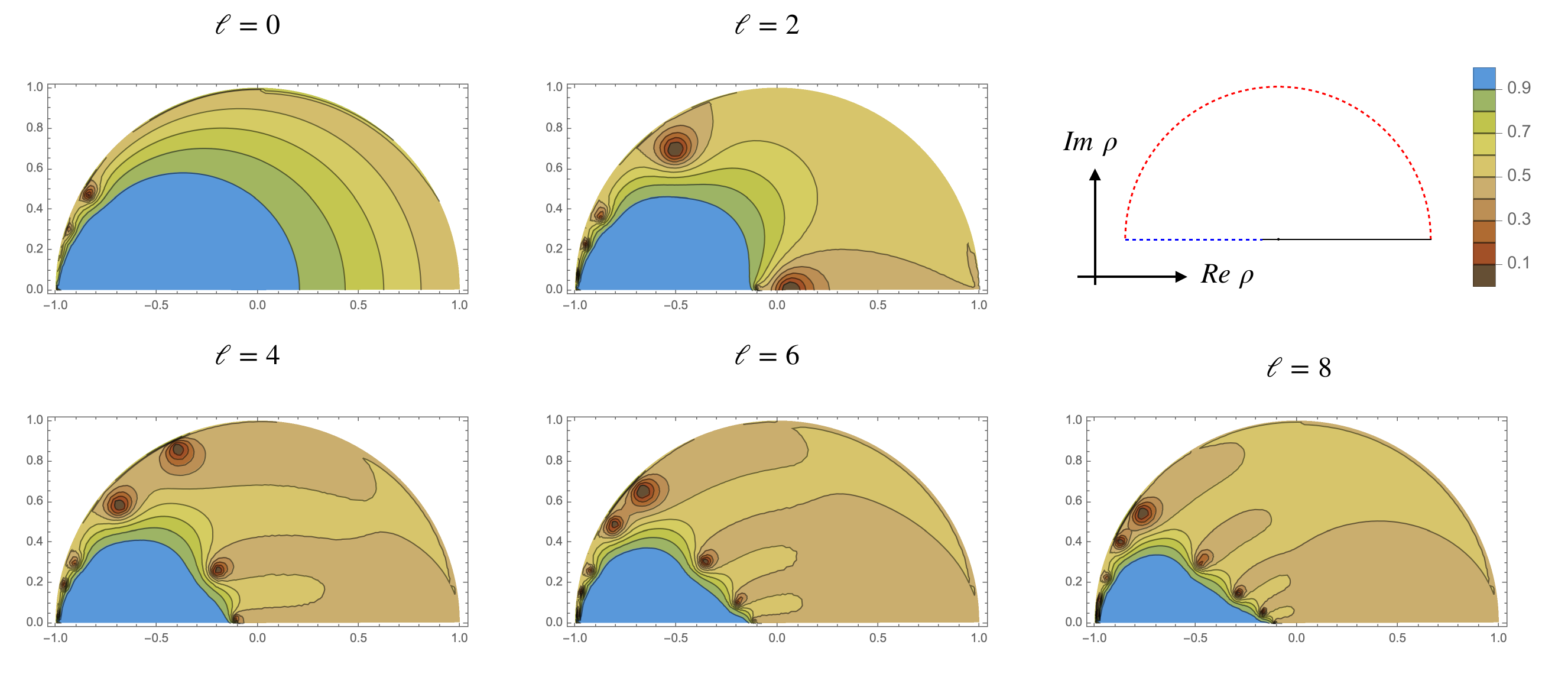}
    \caption{Projection of the partial waves via \( \exp(-\log 2 / |S_\ell(\rho_s)|) \) for \( \ell = 0 \) to \( 8 \), in the upper-half complex \( \rho \)-plane. The red semicircle maps the right-hand cut \( s > 4 \); the blue segment corresponds to the left-hand cut. Resonances appear as brown regions. We use \( p = 20/3 \) for the \( \rho \)-map, and the exponential transformation enhances visibility.}
    \label{fig:half_moon_plots}
\end{figure*}

In Figure~\ref{fig:half_moon_plots}, we show \( |S_\ell(\rho)| \) in the complex \( \rho \)-plane. 
We plot the function $\xi=\exp(-\log 2/|S_\ell(\rho)|)$, such that $\xi=1/2$ when $S_\ell=1$.
Brown regions highlight resonances where $S_\ell \sim 0$, while the blue regions close to the left-cut correspond to values $S_\ell\gg 1$. For \( \ell \geq 2 \), the partial waves exhibit a dense structure of zeros (discussed in Figure~\ref{fig:chew_frauschi}). Interestingly, the resonances associated with the \( Z_i \) trajectories cluster around the left cut, a feature also seen in supergravity amplitudes~\cite{Guerrieri:2022sod}. The \( \ell = 0 \) partial wave shows heavy resonances but lacks accumulation near the left cut. For \( \ell = 2 \), a distinct zero on the real axis signals the virtual state on trajectory \( A' \).

\subsection{Regge trajectories in the complex plane}

\begin{figure*}
\includegraphics[width=0.9\textwidth]{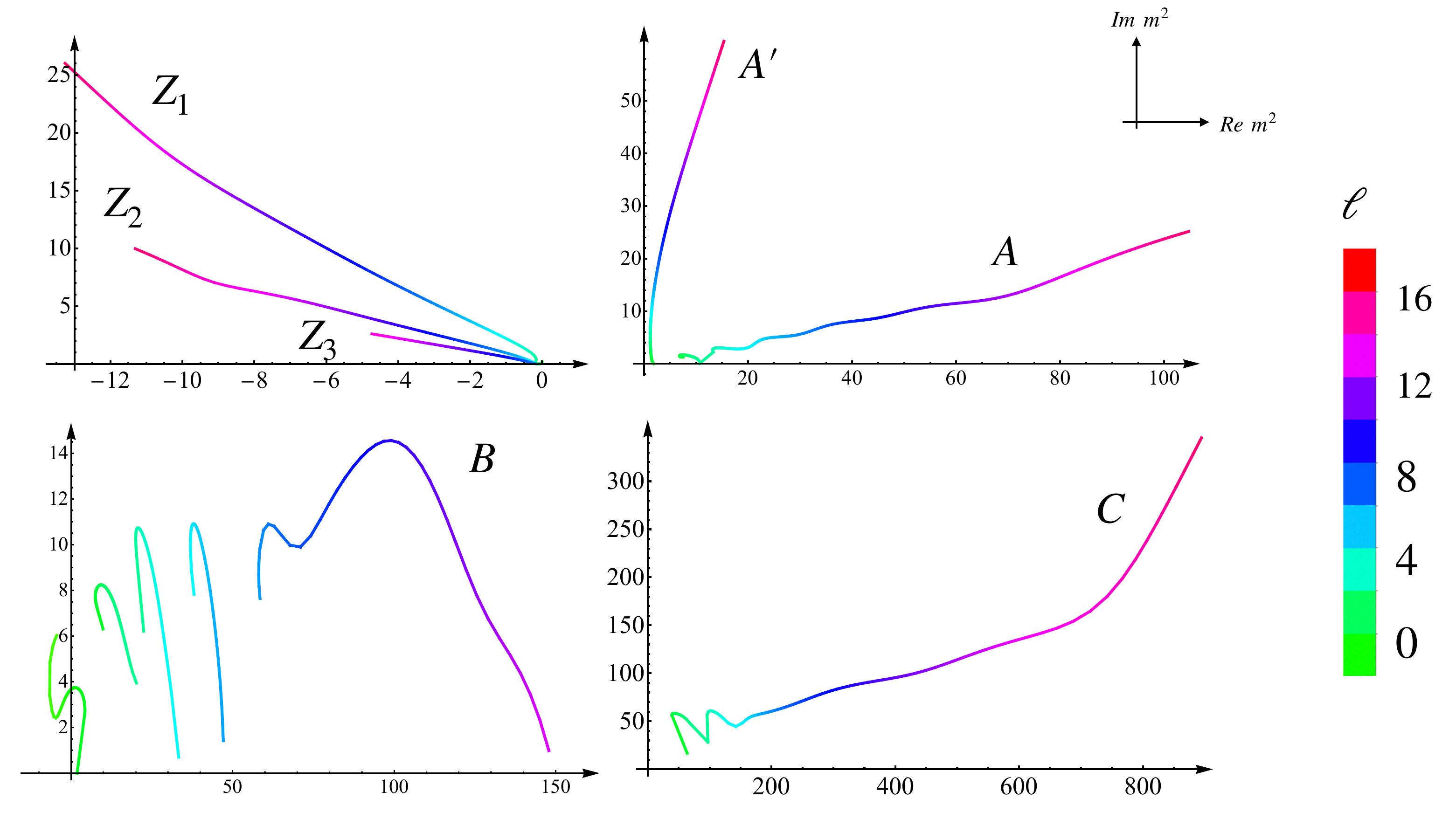}
    \caption{Real and imaginary parts of the squared mass \( m_R^2 \) for the resonances at \( N=20 \), shown as trajectories in the complex \( s \)-plane. Color indicates the value of \( \ell \), with higher spin shown in lighter shades. Some paths cross from the complex plane to the real axis at non-integer values of spin.}
    \label{fig:complex_traj}
\end{figure*}

Figure~\ref{fig:complex_traj} presents a more detailed view of the imaginary-plane projection of the Chew–Frautschi plot shown in Figure~\ref{fig:chew_frauschi}. Here, we display the Regge trajectories in the complex \( s \)-plane, parametrized by spin \( \ell \). The goal is to visualize the analytic continuation of resonances as trajectories in the complex-mass plane.

The Schwarzian trajectories, labeled \( Z_i \), shown in the upper-left corner, appear to be the most stable. These resonances seem to originate near the branch point of the left-hand cut at \( s = 0 \), and move to the left in the complex plane as \( \ell \) increases.

The \( A \) and \( A^\prime \) trajectories are also among the most stable. In particular, the \( A^\prime \) trajectory moves nearly vertically toward the real \( s \)-axis at low spin. The \( A \) trajectory, on the other hand, begins to exhibit oscillatory behavior around \( \ell \sim 6 \), where the imaginary part of \( m_R^2 \) becomes small. These near-real zeros correspond to the sharp \( f_\ell \) peaks observed in the elastic cross-section.

The \( B \) trajectory, associated with the pomeron, is the least stable. While it appears discontinuous in the plot, this is merely an artifact of our visualization: we display only the resonances with \( \Im m_R^2 > 0 \). The oscillatory structure of the \( B \) trajectory has been observed previously in~\cite{Acanfora:2023axz}, where it was associated with complex-spin partial waves exhibiting \( |S_\ell| \gg 1 \) in the physical region. This behavior does not signal a violation of unitarity. As \( N \to \infty \), we observe that the associated resonances tend to migrate from the lower half-plane upward, and the trajectories become increasingly smooth.

\subsection{Asymptotic growth of the Froissart Amplitude}\label{subsec:growatinf}

\begin{figure}[h!]
    \centering    \includegraphics[width=\linewidth]{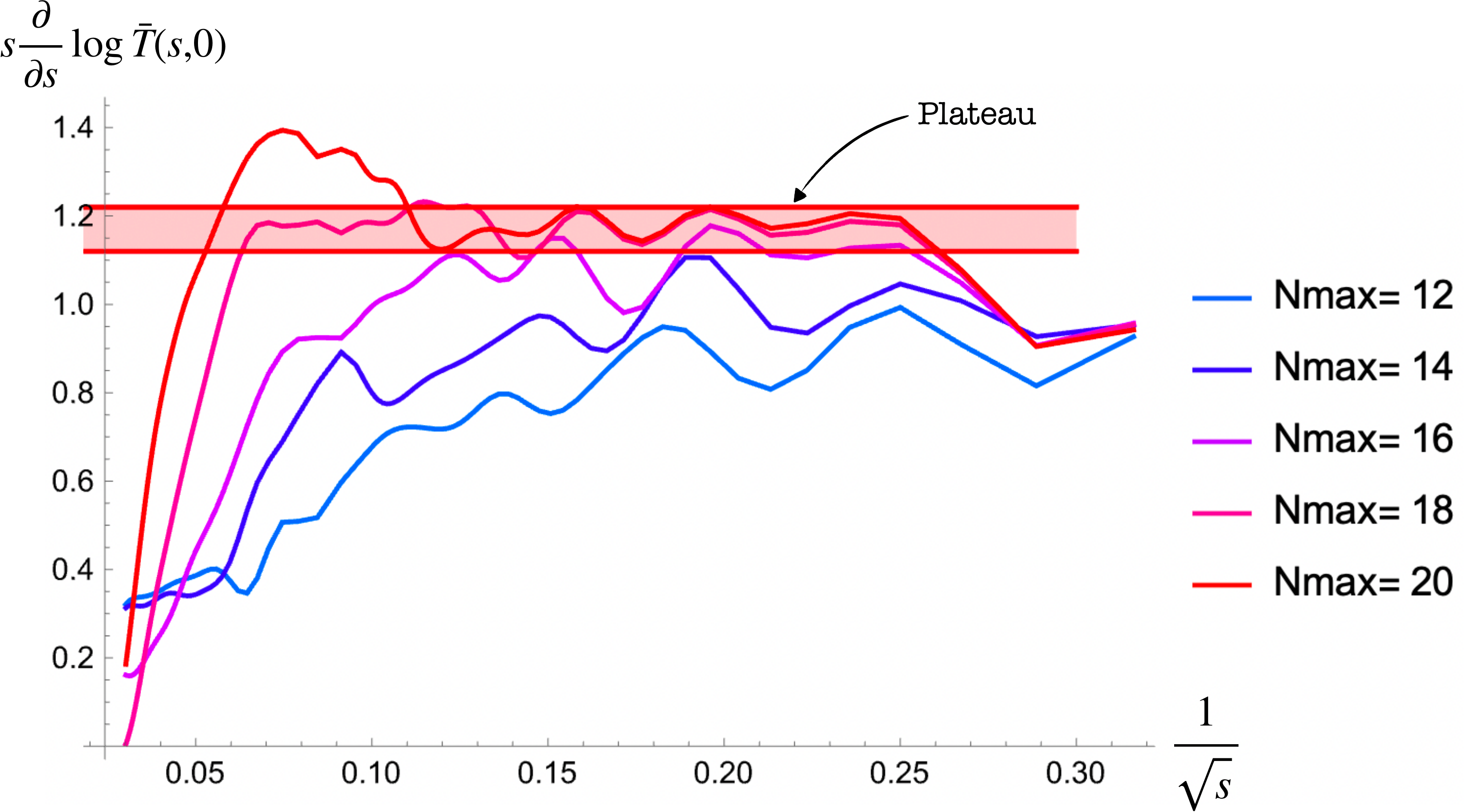}
    \caption{Effective growth exponent \( \alpha_0^\text{eff}(s) \) as a function of \( 1/\sqrt{s} \), computed from a smearing of the amplitude along an arc of radius \( s \) in the complex plane. Different curves correspond to increasing values of \( N_{\text{max}} \). A plateau develops for large enough \( N_{\text{max}} \), indicating approximate Regge behavior at finite energy.}
    \label{fig:effective_growth}
\end{figure}

To probe the asymptotic growth of the Froissart amplitude, we study the effective exponent
\be
\alpha_0^\text{eff}(s) = s \, \frac{d}{ds} \log \overline{|T(s,0)|},
\ee
where \( \overline{|T(s,0)|} \) denotes a smeared version of the amplitude over a circular arc in the complex plane~\cite{Gumus:2023xbs}:
\bea
\overline{|T(s,0)|} &=& \frac{1}{2\pi i} \int_{|v-2|=s} \frac{|T(v,0)|}{v-2} \, dv \nn\\
&=& \frac{1}{2\pi} \int_0^{2\pi} d\theta \, |T(s e^{i\theta}, 0)|.
\eea

Figure~\ref{fig:effective_growth} shows the behavior of \( \alpha_0^\text{eff}(s) \) as a function of \( 1/\sqrt{s} \) for various values of \( N_{\text{max}} \). A notable feature is the emergence of a plateau at intermediate energies for large \( N_{\text{max}} \), indicating that the amplitude mimics power-law growth over a finite energy range.

This plateau, however, is transient. The ansatz used in Eq.~\eqref{eq:foliation} asymptotically tends to a constant as \( s \to \infty \) at fixed \( t \), so \( \alpha_0^\text{eff}(s) \to 0 \) in this limit. Nonetheless, as \( N_{\text{max}} \) increases, the region over which the plateau persists also extends to higher energies.

We interpret this behavior as an effective Reggeization of the amplitude at finite \( N_{\text{max}} \): in the limit \( N_{\text{max}} \to \infty \), one would expect the plateau to extend indefinitely. For \( N_{\text{max}} = 20 \), we estimate the height of the plateau to lie within the range \( 1.12 < \alpha_0^\text{eff} < 1.22 \), in good agreement with the expected pomeron intercept \( \alpha_0 \approx 1.15 \). This growth should eventually unitarize to Froissart-like behavior $\sim s \log^2 s$.

\subsection{Near-forward zeros of the Froissart Amplitude}

A remarkable result by Auberson et al.~\cite{Auberson:1971ru} establishes that amplitudes violating the Pomeranchuk theorem, such as those with cross-sections that grow with energy, must exhibit zeros of the amplitude \(T(s,t)\) near the negative real \(t\)-axis when \(s\) is sufficiently large. When these zeros approach the real \(t\)-axis closely enough, they give rise to observable features in the diffractive cone, notably the appearance of diffractive minima.

\begin{figure}[t]
    \centering
\includegraphics[angle=0, width=0.45\textwidth]{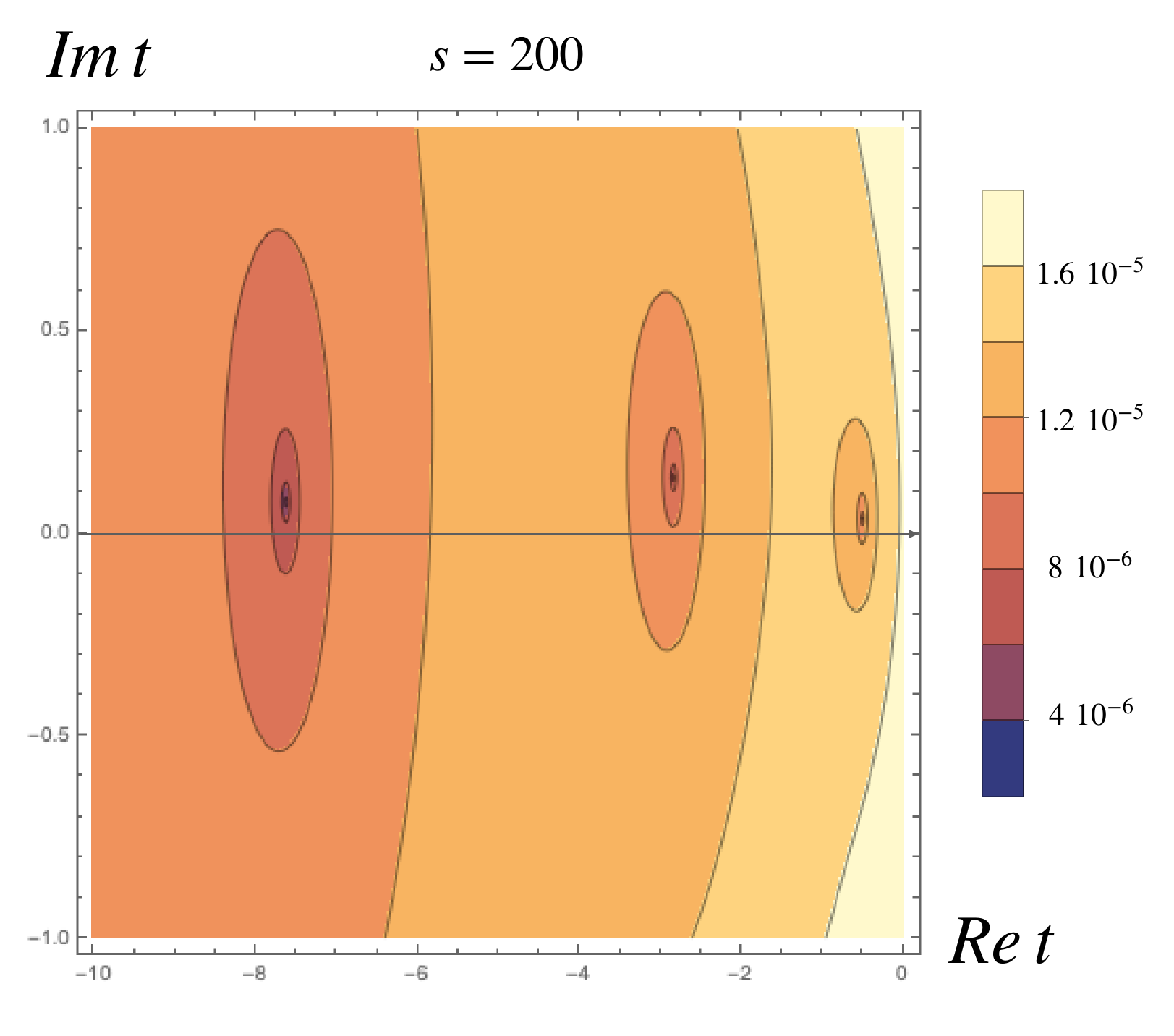}
    \caption{Plot of $\log|T(s=200,t)|/16 s^2$ in the complex $t$-plane. It is evident the presence of near-forward zeros, as anticipated by \cite{Auberson:1971ru}. These zeros are close enough to the negative real axis $t < 0$ and are responsible for the diffractive minima observed in Figure \ref{fig:diffractive_cone}.}
    \label{fig:zeros}
\end{figure}

\begin{figure*}[h!]
    \centering
    \includegraphics[width=0.8\textwidth]{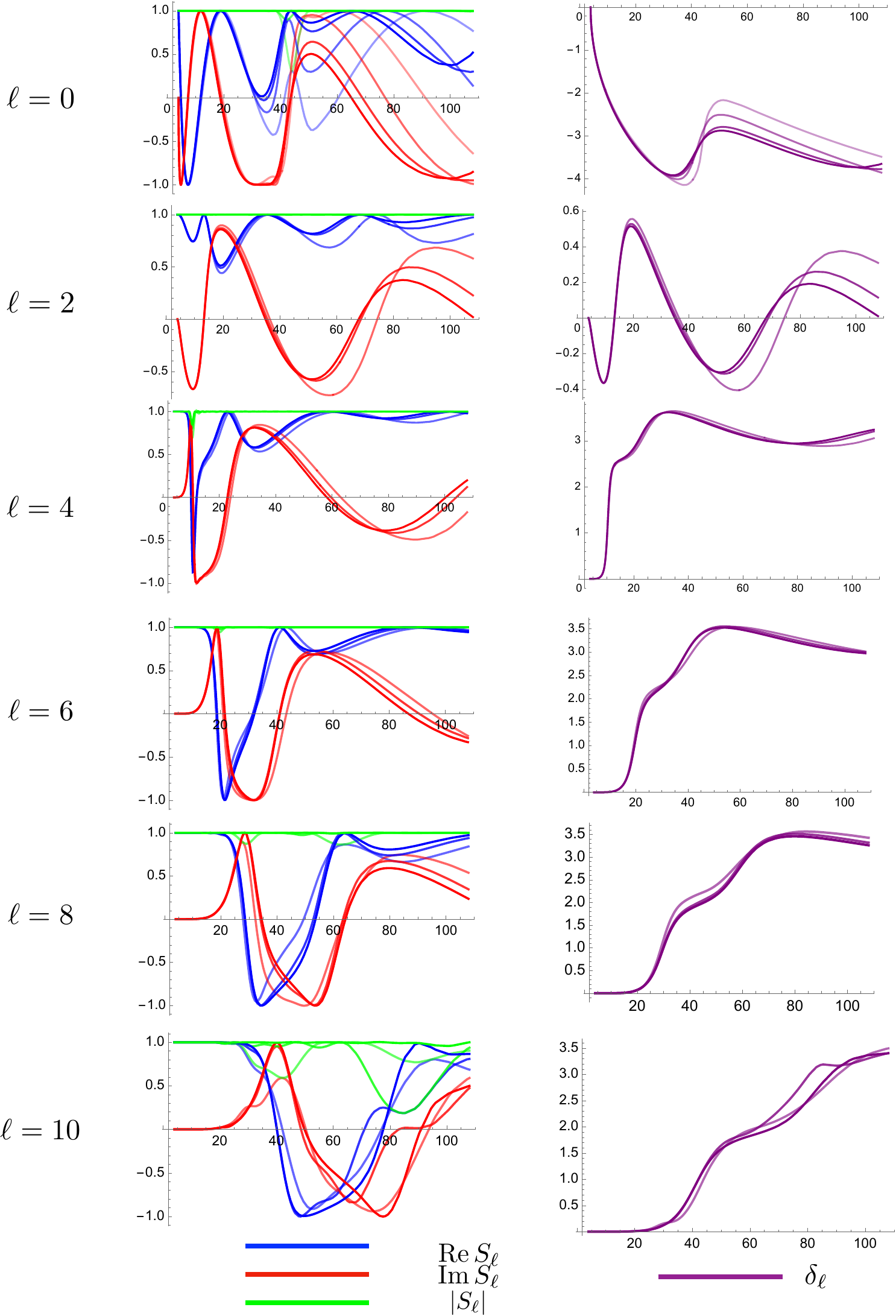}
    \caption{Partial waves and phase shifts of the Froissart Amplitude for $N=16,18,20$ (in opacity gradient) up to spin $\ell=10$.}
    \label{fig:all_partial_waves}
\end{figure*}

In the case of the Froissart amplitude, we find that the diffractive minima indeed stem from such zeros. See Figure \ref{fig:zeros}.

\section{Fitting function for PP scattering and the integrated cross-section}
In order to compare with experimental data and check our analytical and numerical bound we need to parametrize the $pp$ and $p \bar{p}$ scattering total cross-sections. We condense here the results presented in \cite{ParticleDataGroup:2016lqr} for the $pp$ ($p\bar{p}$) cross-section:
\begin{align}
    \sigma_{tot}^{\pm}(s)=& H \log^2\left(\frac{s}{s_M}\right)+ P+ \nonumber \\ &R_1 \left(\frac{s}{s_M}\right)^{-\eta_1}\mp R_2 \left(\frac{s}{s_M}\right)^{-\eta_2}
\end{align}
where $H$ is the coefficient of the Froissart term and controls the growth at infinity. $R_i$ are the intensities of the secondary Regge poles and $P$ is called the Pomeranchuk constant. Finally $s_M=(2m_p+M)^2$, where $m_p$ is the mass of the proton and $M$ is the scale for the ``universal rise of the cross-section".
The parameters take the following values
\begin{align}
    M&=2.1206\pm 0.0094~\text{GeV} \quad H=.2720\pm0.0024~\text{mb}  \nonumber \\ \eta_1&=0.4473\pm0.0077 \quad \eta_2=0.5486\pm0.0049 \nonumber \\
    R_1&=13.07\pm 0.17 ~\text{mb} \quad R_2=7.394 \pm 0.081 ~\text{mb} \nonumber \\ P&=34.41\pm 0.13 ~\text{mb}
\end{align}

\section{Bound on the cross-section assuming a generic kernel.} 
\label{sec:bound_generic_observables}

The choice of the kernel in the definition of the integrated cross-section in Eq.~\eqref{eq:integrated_cross_section} is motivated by both physical and historical considerations.

More generally, under mild assumptions, it is possible to derive analytic bounds on any linear functional of the total cross-section defined via a test function \( \phi \) with compact support \( [a, b] \subset [4, \infty) \),
\be
\mathcal{O}[\phi] = \int\limits_4^\infty \sigma_\text{tot}(s)\, \phi(s)\, ds \equiv \int\limits_a^b \sigma_\text{tot}(s)\, \phi(s)\, ds,
\ee
provided that \( \phi \) is bounded, i.e., \( \phi(s) \leq P \) for all \( s \in [a,b] \).\footnote{
Maximizing the cross-section at a point corresponds to choosing a delta-function kernel. Such a choice lies outside the class of admissible test functions considered here and is not covered by the analytic bounds discussed in this work. In general we do not expect a bound in this case, see Appendix \ref{sec:divcross}.
} The case \( a = 4 \) requires separate treatment. In this instance, it is convenient to define
\[
\phi(s) = (s - 4)^{\tfrac{d}{2} - 1} \, \widetilde{\phi}(s),
\]
where \( \widetilde{\phi}(s) \) is regular as \( s \to 4 \). 

In the remainder of this appendix, we derive the bound for \( a > 4 \), and leave to the intrepid reader who has followed us this far the task of deriving the corresponding bound for \( a = 4 \).
Using the partial wave expansion~\eqref{eq:cross-section}, we have
\bea
&&\mathcal{O}[\phi] = \int_a^b ds\, \phi(s) \sum_{\ell=0}^\infty n_\ell^{(d)}\, \frac{1 - \Re S_\ell(s)}{(s - 4)^{\tfrac{d}{2} - 1}}\leq \nn\\
&& M \frac{2^d \pi^{\tfrac{d}{2} - 1}}{\Gamma(d/2)} (L{-}1)_{d-2}
+ P \int\limits_a^b ds\, \sum_{\ell=L}^\infty n_\ell^{(d)}\, \frac{1 {-} \Re S_\ell(s)}{(s - 4)^{\tfrac{d}{2} - 1}}, \nn
\eea
where
\be
M = \int\limits_a^b ds\, \frac{\phi(s)}{(s - 4)^{\tfrac{d}{2} - 1}} < \infty.
\ee
To control the contribution from high-spin partial waves, we invoke the definition of the dispersive coefficients \( c_{2k}(t_0) \). After applying a series of inequalities, we arrive at
\be
c_{2k}(t_0) \geq \frac{\mathcal{N}_d}{2^k \pi} \, \mathcal{K}_L(a, b, t_0)
\int\limits_a^b ds\, \sum_{\ell = L}^\infty \frac{n_\ell^{(d)} (1 - \Re S_\ell(s))}{(s - 4)^{\tfrac{d}{2} - 1}},
\ee
with the function \( \mathcal{K}_L \) given by
\be
\mathcal{K}_L(a, b, t_0) = P_L\left(1 + \frac{2 t_0}{b - 4} \right)
\frac{\sqrt{a(a - 4)}}{\left( b - 2 + \tfrac{t_0}{2} \right)^{2k + 1}}.
\ee

Combining the results, we obtain a rigorous analytic bound on the observable \( \mathcal{O}[\phi] \),
\be
\mathcal{O}[\phi] \leq
M \frac{2^d \pi^{\tfrac{d}{2} - 1}}{\Gamma(d/2)} (L{-}1)_{d-2}
+ \frac{2^k \pi P\, c_k(t_0)}{\mathcal{N}_d \, \mathcal{K}_L(a, b, t_0)}.
\ee

\end{appendix}

\bibliography{dual_bootstrap}

\end{document}